\documentclass[12pt,preprint]{aastex}

\shorttitle{Variable Stars in NGC~6496}
\shortauthors{Abbas et al.}

\begin{document}

% Moe added this 2013Oct15
\def\citeapos#1{\citeauthor{#1}'s (\citeyear{#1})}


\title{Variable Stars in Metal-Rich Globular Clusters. IV. Long Period
Variables in NGC~6496}


\author{Mohamad A. Abbas\altaffilmark{1}} \affil{Astronomisches
Rechen-Institut, Zentrum f\"{u}r Astronomie der Universit\"{a}t
Heidelberg, M\"{o}nchhofstr. 12--14, D-69120 Heidelberg, Germany}
\email{mabbas@ari.uni-heidelberg.de}

\author{Andrew C. Layden\altaffilmark{2}, Katherine A. Guldenschuh}
\affil{Physics \& Astronomy Department, Bowling Green State
University, Bowling Green, OH 43403, USA}
\email{laydena@bgsu.edu}

\author{D. E. Reichart, K. M. Ivarsen, J. B. Haislip, M. C. Nysewander,
A. P. LaCluyze}
\affil{Department of Physics \& Astronomy, University of North
Carolina, Chapel Hill, NC, 27599, USA}
%\email{aastex-help@aas.org}

\and

\author{Douglas L. Welch\altaffilmark{2}}
\affil{Department of Physics \& Astronomy, McMaster University,
Hamilton, Ontario, L8S 4M1, Canada}


\altaffiltext{1}{Member of IMPRS for Astronomy \& Cosmic Physics at
the University of Heidelberg and of the Heidelberg Graduate School for
Fundamental Physics.  Graduate of Bowling Green State University.}

\altaffiltext{2}{Visiting Astronomer, Cerro Tololo Inter-American Observatory.
CTIO is operated by AURA, Inc.\ under contract to the National Science
Foundation.}

%\altaffiltext{2}{Society of Fellows, Harvard University.}
%\altaffiltext{3}{present address: Center for Astrophysics,
%    60 Garden Street, Cambridge, MA 02138}


\begin{abstract}
We present $VI$-band photometry for stars in the metal-rich globular
cluster NGC~6496.  Our time-series data were cadenced to search for
long period variables (LPVs) over a span of nearly two years, and our
variability search yielded the discovery of thirteen new variable stars, of
which six are LPVs, two are suspected LPVs, and five are short period
eclipsing binaries.  An additional star was found in the ASAS
database, and we clarify its type and period.  We argue that all of
the eclipsing binaries are field stars, while 5-6 of the LPVs are
members of NGC~6496.  
%The period-luminosity distribution of these LPVs
%agree well with those found in the Large Magellanic Cloud and 47
%Tucanae, and with theoretical pulsation models for these systems.
We compare the period-luminosity distribution of these LPVs with those
of LPVs in the Large Magellanic Cloud and 47 Tucanae, and with
theoretical pulsation models.  We also present a $VI$ color magnitude
diagram, display the evolutionary states of the variables, and match
isochrones to determine a reddening of $E(B-V) = 0.21 \pm 0.02$ mag
and apparent distance modulus of $15.60 \pm 0.15$ mag.
%We suggest/recommend
%more searches for LPVs in different globular clusters in order to
%better understand the correlation of the period-luminosity
%distribution with other parameters as well as to improve the pulsation
%models.
\end{abstract}


\keywords{variable stars: general --- globular clusters: individual
(NGC 6496) --- stars: horizontal-branch }



\section{Introduction}  \label{sec_intro}

Models of stellar structure and evolution for low-mass stars have been
developed and improved using observational data of globular clusters
(e.g., \citealt{sweigar1998,piotto2007}). For instance, observing a
globular cluster in two or more bands allows an observer to obtain the
color-magnitude diagram (CMD). Comparing the CMD to isochrones yields
the locations and shapes of the principle sequences, which can be used
to refine the theoretical treatment of stellar interiors.
Additionally, matching the observed CMD to isochrones provides
information about the metallicity and age of the observed globular
cluster.  In a similar way, comparisons of observed stellar pulsation
properties (periods, amplitudes, surface velocities, etc.) with the
predictions of theoretical models of stellar pulsation can provide
additional constraints on the properties of stellar interiors and
evolution (e.g., \citealt{smith1995,olivwood05,lebwood05}).
%However, questions about stellar pulsation models and stellar 
%evolution can not be answered using the color-magnitude diagrams 
%and isochrones alone. 

To obtain such data, time series observations are needed to detect
variable stars and follow their properties over multiple pulsation
cycles.
%study their properties which in return
%improve our understanding of stellar pulsation and evolution
%\citep{smith1995}.
Previous variability studies of globular clusters focused on short
period variable stars such as RR Lyrae stars, SX Phoenicis
stars, etc., as the time series observations could be acquired during
a classically-scheduled observing run.
%over a short period of time . Observing these rapid variables
%repeatedly over few days is enough to obtain well sampled light curves
%and periods. 
It is more challenging and time consuming to
detect long period variables (LPVs) as they have periods ranging from
30-1000 days, but the new generation or robotic telescopes have
enabled multi-year campaigns with short and regular observing cadences.  
%the latter stars are slow
%variables. The stellar pulsation of these stars is detectable only
%over a long period of time (e.g., $\textgreater 100$ days).
%Depending on the period and amplitude of their pulsation, 

Typically, LPVs are divided into Mira, Semiregular, and Irregular
classes, forming a continuum in descending order of period length,
amplitude size, and regularity of cyclicity.  
%Long period variables are located on the asymptotic giant branch of 
%the color magnitude diagrams and have periods in the 30--1000 days
%range. 
These stars are in their last stage of stellar evolution, on the
asymptotic giant branch (AGB), and are forming dust and ejecting gas
while pulsating. Understanding the pulsation and properties of
these stars
%in their last stage can thus 
can help us improve our understanding of theoretical stellar pulsation
and evolution along with the process of chemical enrichment of the
interstellar medium.

As part of a survey for RR Lyrae stars in metal rich globular
clusters \citep{layden99, layden03, baker07},
%clusters (\citealt{layden99,layden03,baker07}; Papers I, II, and
%III, respectively), 
%and for the irregular series, 
we present multi-band, multi-epoch observations obtained for the
metal-rich globular cluster NGC~6496 (C1755-442).  This cluster is
located
%$($\alpha_{2000} = 17^{h}59^{m}02^{s}.0$,
%$\delta_{2000}=-44^\circ15\arcmin54\arcsec$) 
at galactic coordinates ($l$, $b$) = (348.03\degr, --10.01\degr), at a
distance $\sim$11.3 kpc from the Sun.  The moderate latitude comes
with a modest level of foreground reddening, $E(B-V) = 0.15$ mag, and
field-star contamination.  The cluster has a relatively metal-rich
composition of [Fe/H] = --0.46 \citep{harris96}, leading
% some to propose that it is
\citet{zinn85} to list it as a member of the galactic thick disk
population, though \citet{richtler94} provided counter evidence.  The
cluster has an open, uncroweded nature (structural parameters include
core, half-mass, and tidal radii of 0.95, 1.02, and 4.8 arcmin,
respectively) making NGC~6496 a good candidate for CCD photometry with
a telescope of modest plate scale.  However, the faint integrated
light of the cluster after conversion to absolute magnitude, $M_V =
-7.2$ mag \citep{harris96}, and its consequent low integrated stellar
mass indicate
% a cluster with relatively few member stars, and thus 
that a small population of variable stars is to be expected.  Previous
searches for variable stars in the cluster found no candidates
\citep{clement01}.  In particular, the photographic survey by
\citet{fourcade66} yielded no clear variable stars.  A lack of RR
Lyrae variables in NGC~6496 is no surprise given the cluster's
relatively high metallicity and red-clump horizontal branch morphology
\citep{armand88,richtler94,saraj94}, but the high metallicity favors
the formation of long period variables (for example, \citet{frogel98}
showed that luminous LPVs are found only in clusters with [Fe/H] $>
-1.0$), and so a modern search seems warranted.

In this paper, we present new time-series images in Section
\ref{sec_obsreds}, our photometric analysis of these data in Sec.
\ref{sec_daophot}, and the resulting color-magnitude diagrams in Sec.
\ref{sec_cmd}.  We discuss the detection, properties and membership
likelihood of the variable stars in Sec. \ref{sec_varstar}, and
compare them with similar variables in the Large Magellanic Cloud and
the metal-rich globular cluster 47 Tucanae.  We summarize our findings
in Sec. \ref{sec_summary}.

% Moe added references:
%\bibitem[Sweigart \& Catelan(1998)]{sweigar1998} Sweigart, A.~V., \&
%Catelan, M.\ 1998, \apjl, 501, L63

%\bibitem[Piotto et al.(2007)]{piotto2007} Piotto, G., Bedin, L.~R., Anderson, J., et al.\ 2007, \apjl, 661, L53

%\bibitem[Smith(1995)]{smith1995} Smith, H.~A.\ 1995, RR Lyrae Stars. Cambridge Univ. Press, Cambridge

%Harris 1996 data (downloaded Jan 2014 from website)
%RA, Dec =        17 59 03.68  -44 15 57.4  
% l = 348.03  b =  -10.01
%
% [Fe/H] wt  E(B-V) V_HB (m-M)V V_t   M_V,t   
%  -0.46  3   0.15 16.47 15.74  8.54  -7.20 
%  U-B   B-V   V-R   V-I  spt   ellip
% 0.45  0.98              G4    0.16
%
%     v_r   +/-    v_LSR    sig_v  +/-    c        r_c   r_h
%  -112.7   5.7  -108.4                 0.70      0.95  1.02 
%  r_t = r_c * 10^c = 0.95' * 10^0.70 = 4.76'
%    mu_V   rho_0 lg(tc) lg(th)
%   20.17   1.99   8.94  9.04

\section{Observations and Reductions}  \label{sec_obsreds}

As part of the original survey for RR Lyrae stars in metal-rich
globular clusters, we obtained time-series images of NGC~6496 using
the direct CCD camera on the 0.9-m telescope at Cerro Tololo
Inter-American Observatory (CTIO) during two runs in May and June of
1996 having 1 and 7 usable nights, respectively. The Tek\#3 2048 CCD,
operated with a gain of 2.4 electrons/adu and a readout noise of 3.7
electrons, provided a $13.5\arcmin$ field of view with $0.4\arcsec$
pixels.  We used filters matched to the CCD to reproduce the Johnson
$V$ and Kron-Cousins $I$ bandpasses.  The raw images were processed
using the Image Reduction and Analysis Facility (IRAF)\footnote{IRAF
is distributed by the National Optical Astronomy Observatories, which
are operated by the Association of Universities for Research in
Astronomy, Inc. under cooperative agreement with the National Science
Foundation.} to perform the standard procedures for overscan
subtraction and bias correction, and by using twilight sky frames to
flat-field the images.

In a typical pointing toward the cluster, we obtained a short exposure
(25--120 s) $VI$ frame pair and a long exposure (250--350 s) $VI$
pair.
%This provided two independent magnitude estimates of the HB
%stars at each observational epoch, and extended the dynamical range of
%the observations.  
Such pointings were obtained 1--2 times each night, with time
intervals between pointings of at least five hours.  In total, we
obtained 48 images in twelve pointings (independent photometric
epochs) toward NGC~6496.  The seeing varied between 1.0$\arcsec$ and
4.5$\arcsec$ (1.9$\arcsec$ median) fwhm, with 19\% of the images
having seeing greater than 2.5$\arcsec$.
%In images taken during good seeing, a radial gradient in the quality of
%the stellar profiles was evident, and in some images, the stars near
%the CCD corners were too far out of focus to yield reliable
%photometry.
Preliminary photometry using the CTIO images indicated the presence of
at least five slow, red variables, and at least five stars with
variations on shorter timescales.  However, the sampling was
insufficient to determine the nature of these variables.

Additional observations were obtained using the 0.41-m Panchromatic
Robotic Optical Monitoring and Polarimetry Telescope number 4
(PROMPT4; \citealt{reichart05}).  This telescope is equipped with a
1k$\times$1k Apogee CCD camera \citep{nysewander09} with a 10$\arcmin$
field of view and a 0.59$\arcsec$ pixel size \citep{reichart05}. The
CCD has a gain of 1.5 electrons/adu and readout noise of 13.5
electrons.  On $\sim$63 nights that span almost two years (February
2009 through October 2010) and that are each separated by 1-2 weeks,
we obtained images of NGC~6496 through standard $V$ and $I$ filters.
On each night, we obtained four 80 s images in $V$, three 40 s images
in $I$, and three 10 s images in $I$ (hereafter referred to as $I_L$
and $I_S$, for ``long'' and ``short,'' respectively).  The images were
taken in the order ``$V$, $I_L$, $I_S$; $V$, $I_L$, $I_S$; $V$, $I_L$,
$I_S$; $V$'' so that small tracking errors would shift stars across
several pixels, providing {\it de facto} dithering that would later
allow us to reject bad pixels and cosmic rays.  The long and short
exposure $I$ images allowed us to maximize dynamic range in the
resulting photometry, and ensure that bright LPVs were not saturated
when observed at maximum light.  The seeing varied between 1.2 and
3.5$\arcsec$ fwhm with a median of 1.9$\arcsec$, and with 13\% of
the 188 individual P4 images having seeing greater than 2.5$\arcsec$.

%DAOPHOTv/fwhm_stats.doc   2.1 to 5.9 pix  med 3.5
%DAOPHOTis/fwhm_stats.doc  2.1 to 4.6 pix  med 3.1
% *0.59''/pix =            1.2    3.5          1.9

\begin{figure}
\epsscale{1.0}
\plotone{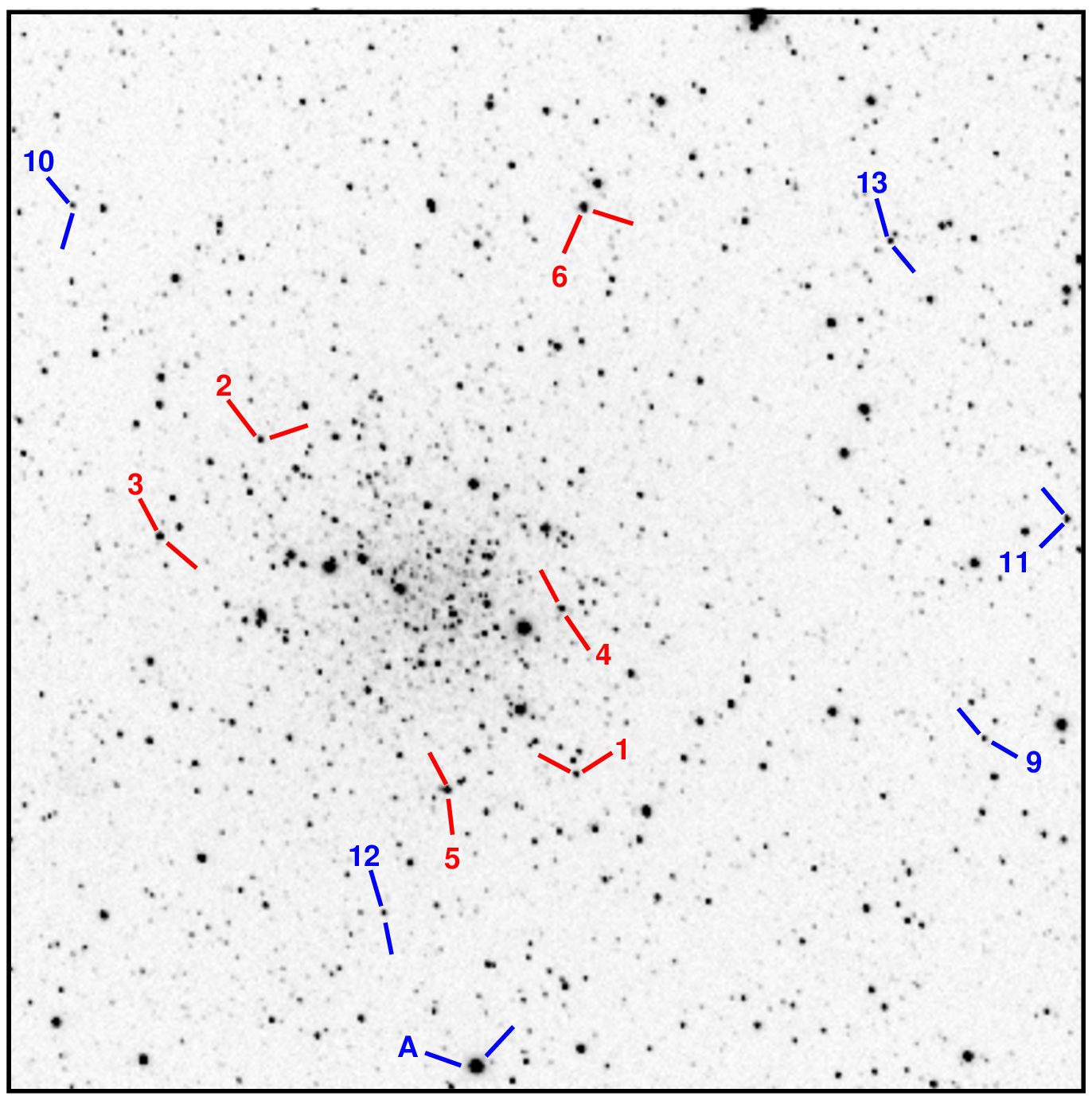}
%\plotone{frame2.eps}
\caption{The combined $V$-band master image of NGC~6496 taken with
PROMPT4. North is up and East is to the left.  The axes of the $10 \times 10$ arcmin field of view 
%are marked in pixels. 
go from $X_{pix} = 1$ to 1024 pix from left to right, and from
$Y_{pix} = 1$ to 1024 pix from bottom to top.  The uncrowded nature of
the cluster helped us obtain good photometry even near the cluster
center.  Variable stars are marked with their assigned numbers (e.g.,
"1" indicates "V1"), while ``A'' marks the variable star ASAS
175901-4411.5.  In the electronic edition, red and blue symbols mark
long and short period stars, respectively.
\label{masterimage}}
\end{figure}

The raw cluster images and suitable calibration images from each
observing night were retrieved from the PROMPT website and processed
using IRAF 
% the Image Reduction and Analysis Facility (IRAF; \citealt{tody1986}) 
packages to correct for bias patterns, dark
current, and flatness-of-field variations.  After inspecting the set
of four processed $V$ images from each pointing and rejecting any poor-quality ones, the
remaining images where shifted and combined to produce a single, high
signal-to-noise (S/N) $V$ image that is devoid of bad pixels and
cosmic rays.  Similar steps were followed separately for the three
$I_L$ images and for the three $I_S$ images, yielding three final
images for each observing night.  One such image, the combined $V$
frame taken on 2009-08-20, is shown in Figure \ref{masterimage}.  Note
the open nature of the cluster, and the significant field star
population.
Together, the CTIO and PROMPT images provided up to 77 independent photometric epochs in each band for each variable star in our data set.

%% In a manner similar to \objectname authors can provide links to dataset
%% hosted at participating data centers via the \dataset{} command.  The
%% second curly bracket argument is printed in the text while the first
%% parentheses argument serves as the valid data set identifier.  Large
%% lists of data set are best provided in a table (see Table 3 for an example).
%% Valid data set identifiers should be obtained from the data center that
%% is currently hosting the data.
%%
%% Note that AASTeX interprets everything between the curly braces in the 
%% macro as regular text, so any special characters, e.g. "#" or "_," must be 
%% preceded by a backslash. Otherwise, you will get a LaTeX error when you 
%% compile your manuscript.  Special characters do not 
%% need to be escaped in the optional, square-bracket argument.

%% In this section, we use  the \subsection command to set off
%% a subsection.  \footnote is used to insert a footnote to the text.


%% This section also includes several of the displayed math environments
%% mentioned in the Author Guide.

\section{DAOPHOT Photometry}  \label{sec_daophot}

Instrumental photometry of the CTIO 0.9-m images was performed using the
DAOPHOT~II \citep{stetson87} and ALLFRAME \citep{stetson94} packages
following the procedures described in \citep{baker07}.  Hereafter, we
refer to this as the ``CTIO'' photometry set.

%We transformed this instrumental photometry onto the standard $VI$
%system using uncrowded stars in the $VI$ photometry of
%\citep{armand88} as secondary standards.

These packages were used in a similar way to do photometry on the
PROMPT4 images (hereafter referred to as the ``P4'' photometry set).  The
FIND and PHOT routines of DAOPHOT~II were used to detect and provide
initial aperture photometry of objects above a threshold on each
image.  We then used PICK to select $\sim$160 bright stars located
away from the cluster center to determine the stellar point-spread
function (PSF), and the PSF routine was used to combine all the PSF
stars' profiles into a single PSF model for each image.  The ALLSTAR
program then fit the PSF model to all of the stars in each processed
image and gives an improved estimation of the stars' locations and
magnitudes.  An automatic, iterative procedure was used to subtract
the light of stars located near the PSF stars and build an improved
PSF, which was run again through ALLSTAR to further improve the
photometry for that image.

We then selected the best ten $V$ images and combined them using
MONTAGE to create a deep image with high S/N that can be used as a
reference image. The DAOPHOT and ALLSTAR procedures were then applied
to the MONTAGE image to get a master reference list that contains the
locations and average magnitudes of all sources found in the MONTAGE.
Similarly, MONTAGE images and master star lists were created from the
$I_L$ and $I_S$ image sets.

The DAOMASTER program was used to match the same stars in different
images.  Using the master list of stellar positions, the ALLFRAME
program was used (separately for the $V$, $I_L$ and $I_S$ sets) to
obtain final instrumental photometry for objects on each frame.  The
systematic use of a single master star list reduces the confusion in
crowded environments that can lead to false detections of variable
stars.  Finally, the photometry files, one from each image, were combined
using DAOMASTER to provide a list of instrumental averaged magnitudes
(the DAOPHOT ``.mag'' file) and a file with instrumental time-series
photometry (the DAOPHOT ``.raw'' file) for each star found in the
MONTAGE image (again, separately for $V$, $I_L$, and $I_S$).

The P4 camera was removed for service and replaced around 2009 July
01.  Images taken before and after this date had slightly different
rotations with respect to the equatorial coordinate system, and we
found our ALLFRAME results were better when we treated these groups of
images separately.  We used DAOMASTER to merge these data sets as
described above, and then to match occurrences of the same star in $V$,
$I_L$, and $I_S$ to compile a single list of averaged star magnitudes.

\begin{figure}
\epsscale{1.0}
\plotone{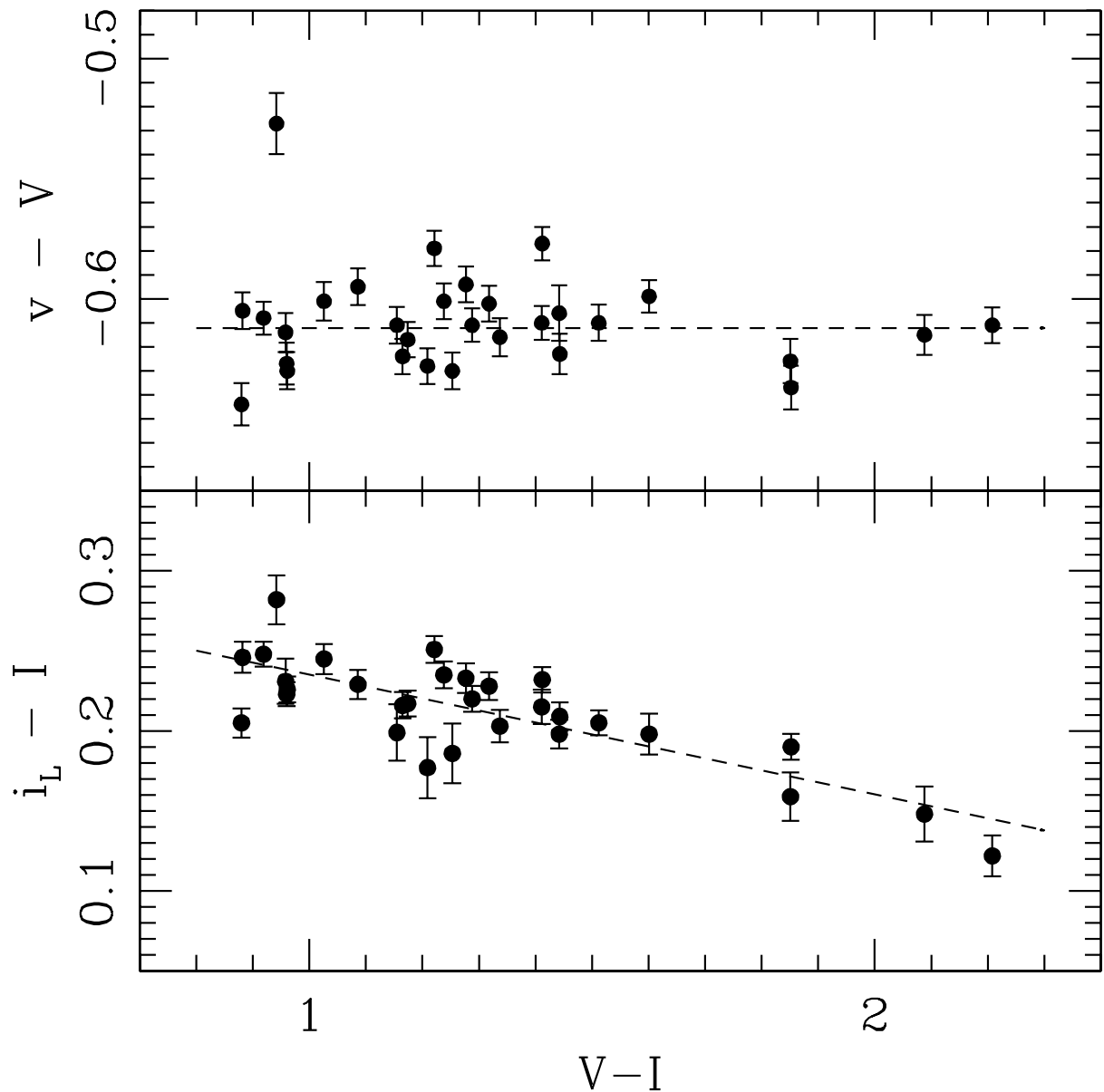}
%\plotone{stetson_comp2014.eps}
\caption{The data used to transform our P4 photometry to the
standard system.  (Top) the difference between our instrumental $v$
magnitudes and \citet{stetson00} $V$ magnitudes, and (bottom) our
$i_L$ and Stetson's $I$ magnitudes, are plotted against Stetson's
color index, $V-I$. The dashed lines correspond to the transformation
coefficients listed in Table \ref{coeff}.
\label{calibstet1}}
\end{figure}

Since all the P4 photometry obtained from DAOPHOT is instrumental, we
used a list of 33 photometric standard stars in NGC~6496 from
\citet{stetson00}\footnote{See
\tt{http://www4.cadc-ccda.hia-iha.nrc-cnrc.gc.ca/community/STETSON/standards/}.}
in common with our P4 data,
to obtain transformation Equations \ref{coeff_v}, \ref{coeff_il} and \ref{coeff_is}:
\begin{equation}  \label{coeff_v}
v - V =  \alpha_1 + \beta_1(V-I)
%v - V =  \alpha_1 - \beta_1(v-i)
\end{equation}
\begin{equation}  \label{coeff_il}
i_L - I =  \alpha_2 + \beta_2(V-I)
%i_L - I =  \alpha_2 - \beta_2(v-i_L)
\end{equation}
\begin{equation}  \label{coeff_is}
i_S - I =  \alpha_3 + \beta_3(V-I)
%i_S - I =  \alpha_3 - \beta_3(v-i_S)
\end{equation}
where $V$ and $I$ represent the standard magnitudes from
\citet{stetson00}, while $v$, $i_L$, and $i_S$ correspond to our
instrumental magnitudes obtained above for the 33 photometric standard
stars. The least-squares fits were obtained, and the resulting
coefficients and their uncertainties ($\epsilon$) are reported in
Table \ref{coeff}, along with the RMS of the points around the best
fit line.  The fits for Equations \ref{coeff_v} and \ref{coeff_il} are
plotted in Figure \ref{calibstet1}.  Plots showing $(v-V)$ and $(i-I)$
versus magnitude and location on the chip showed no significant
trends.  This and the RMS values from Equations \ref{coeff_v} through
\ref{coeff_is} indicate that the zero-point calibrations of our P4
photometry are reliable to better than 0.005 mag.  Similar fits were
performed for the CTIO data (see Table \ref{coeff}), but the RMS
values were larger, so we place higher weight on the P4 data.

With these relations in hand, we transformed our averaged,
instrumental P4 magnitudes onto the Stetson standard $V$ magnitude
system.  For each star, the first step was to calculate the star's
color via
\begin{equation}  \label{coeff_vi}
V - I = \frac{ (v-i_x) - (\alpha_1 - \alpha_x)}{1 + (\beta_1 - \beta_x) },
\end{equation}
where $x = 2$ or 3, for the long and short-exposure $I$ data,
respectively.  Using Equations \ref{coeff_v}, \ref{coeff_il} and
\ref{coeff_is} and the least-squares coefficients found in Table
\ref{coeff}, we computed the star's magnitudes in $V$, $I_L$, and
$I_S$.
% , along with the weighted average of the $I_L$ and $I_S$ magnitudes.  
%We then recompiled the list of P4 data, retaining whichever $I$ magnitude
%($I_L$ or $I_S$) had the smaller error, yielding photometry for 1675
This yielded photometry for 1675 stars, 491 of which areq brighter
than $V = 16.5$ mag, the approximate level of the horizontal branch.
% 689 brighter than V=17.0; 1322 brighter than V=18.  
For 330 stars with $11 < V < 15$ mag, the mean difference, $I_L -
I_S$, was +0.006 mag with an RMS scatter of 0.015 mag; because both
sets of magnitudes were brought successfully to the same standard
system, we computed the weighted mean of the $I_L$ and $I_S$
magnitudes for each star, along with its weighted error.

Similar transformations were applied to the CTIO instrumental
photometry using the coefficients from Table \ref{coeff}, yielding
standard $VI$ photometry for 11,207 stars, 666 of which are brighter
than $V = 16.5$ mag.
%\footnote{Most of the stars added with respect to the P4 data are fainter than the HB, and at 
%large projected radius from the cluster center.  Dominated by fainter field stars, this extra 
%data adds little to our study of the cluster and its variables}.
The mean differences between the CTIO and P4 magnitudes are $+0.004$
mag in both $V$ and $I$, with rms scatters of 0.037 and 0.044 mag,
respectively.  Tests suggest that the larger RMS values here, and for
the CTIO transformation relations reported in Table \ref{coeff},
reflect weak quadratic trends
%systematic zero-point variations
with $X$ and $Y$ position in our CTIO instrumental photometry having
center-to-corner amplitudes of less than 0.1 mag.
%at the level of several hundredths of a magnitude.  
These are probably due to the radial focus variations in the CTIO
0.9-m images described by \citet{baker07}, whose images were acquired
during the same observing runs as the CTIO images of NGC~6496.  Rather
than correct for this effect, we use the calibrated P4 data to
represent the time-averaged photometry of stars in this cluster.

Next, we compare our P4 standard photometry with published photometry
of bright stars ($V < 16.5$ mag).  For 107 stars in common with the
$VRI$ data of \citet{armand88}, we found the mean differences $V_{us}
- V_{A88} = +0.091 \pm 0.003$ mag (RMS = 0.027 mag) and $I_{us} -
I_{A88} = +0.056 \pm 0.005$ mag (RMS = 0.047 mag).  The difference in
both filters showed a small but significant trend with $V-I$ color.
For 43 stars in common with the $BV$ data of \citet{richtler94} and
\citet{richtler95}, we found a mean difference of $V_{us} - V_{R} =
-0.111 \pm 0.006$ mag (RMS = 0.039 mag) with no trend in $B-V$ color.
For 64 stars in common with the $BV$ data of \citet{saraj94}, we found
a mean difference of $V_{us} - V_{SN} = -0.054 \pm 0.006$ mag (RMS =
0.045 mag) with a slight trend in $B-V$ color.
% See /physics1/people/layden/N6496/MOE_ABBAS/fitrot standards2014.doc

Because our data falls near the centroid of these $V$ calibrations,
and because the \citet{stetson00} standards to which our data are tied
used a very large number of observations on many different photometric
nights, we argue that our mean $VI$ photometry is among the best
currently published for NGC~6496.  We present our photometry in Table
\ref{tab_cmd}, where the columns indicate (1) our identification
number, (2) the $X$ and (3) $Y$ pixel coordinates\footnote{These
coordinates correspond to those in Figure \ref{masterimage}, and to
the FITS-format image available at {\tt
http://physics.bgsu.edu/$\sim$layden/publ.htm}, where additional data
products from this study are available.}, (4) the $V$ magnitude and
(5) its uncertainty, (6) the weighted $I$ magnitude and (7) its uncertainty.
Columns 9 and 10 show the mean ``chi'' and ``sharp'' image diagnostics
described in \citet{stetson87}.

%%%%%%%%%%%%%%%%

\section{Color-Magnitude Diagrams}   \label{sec_cmd}

\begin{figure}
\epsscale{1.0}
\plotone{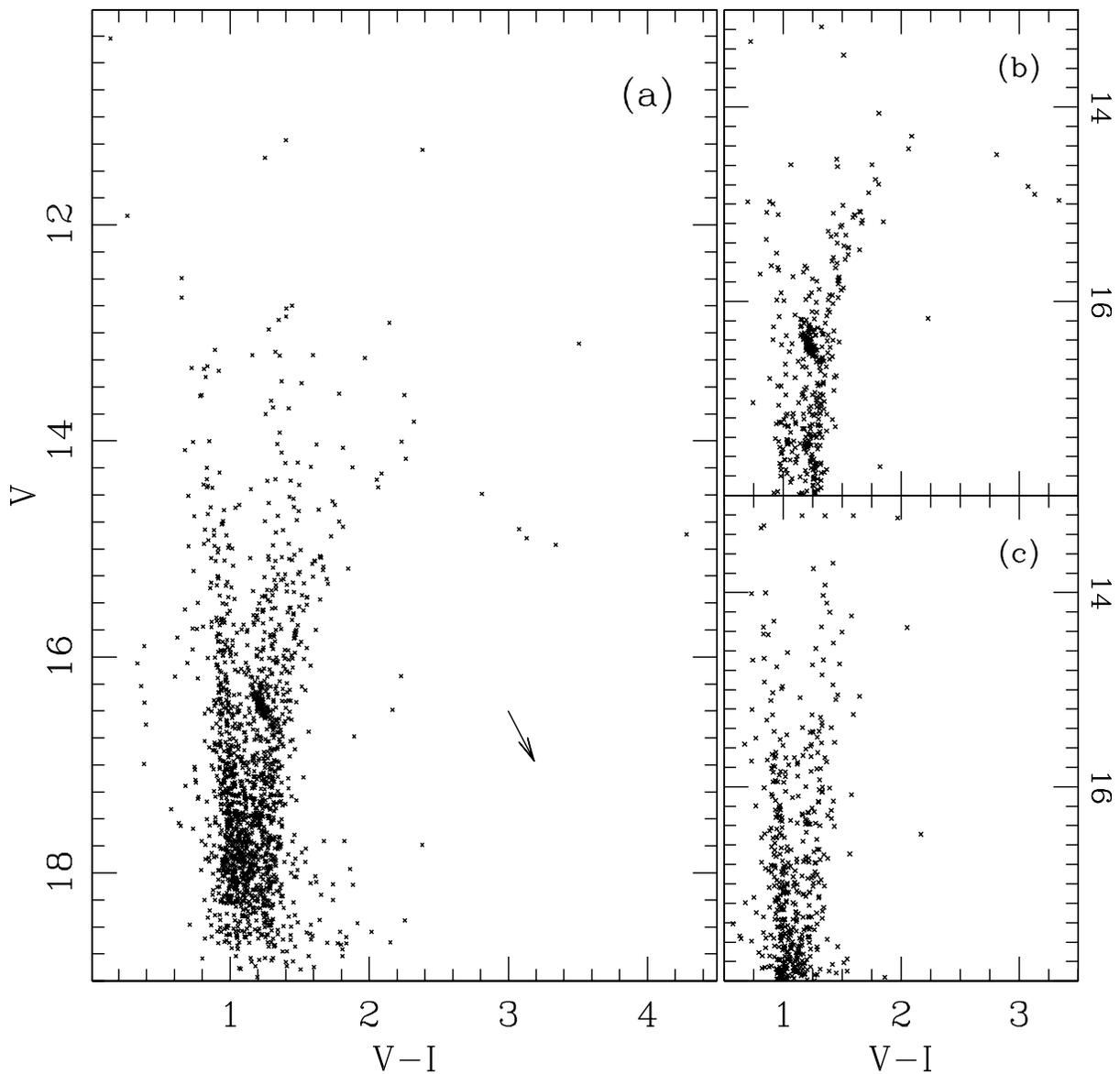}
%\plotone{cmd3.eps}
\caption{Color-magnitude diagrams in ($V-I$, $V$) for NGC~6496.  (a)
All stars in our data set are shown.  The arrow indicates the
reddening vector for $E(B-V) = 0.15$ mag \citep{harris96}.  (b) Only
stars located within 2.5 arcmin of the cluster center are shown,
giving us a clearer depiction of the cluster CMD.  (c) Only stars
located ouside the tidal radius of 4.8 arcmin \citep{harris96} are
shown, representing the field star population toward NGC~6496.
\label{fig_cmd3}}
\end{figure}

We use the calibrated P4 data to plot the ($V-I$, $V$) CMD of NGC~6496
in Figure \ref{fig_cmd3}.  Panel (a) shows all the stars in our data
set, which includes many stars outside the cluster's tidal radius of
4.8 arcmin \citep{harris96}.  The cluster HB is evident at ($V-I$,
$V$) = (1.2, 16.5 ) mag.  It is stretched parallel to the reddening
vector, indicating that some differential reddening ($\lesssim 0.2$
mag) is present across our 10 arcmin field of view.  The differential
reddening also stretches the RGB, lowering its contrast with respect
to the background of field stars noted in Figure \ref{masterimage}.
The sequence of blue stars with $V-I < 1.0$ and many of the redder
stars with $V < 15$ are typical of field stars seen in many studies of
low-latitude globular clusters.

Panel (b) of Figure \ref{fig_cmd3} shows only stars located within 2.5
arcmin of the cluster center (the choice of 2.5 arcmin is an arbitrary
compromise between limiting the field population and maximizing the
cluster population).  The HB is less stretched, indicating less
differential reddening over this more compact region of the sky, and
the RGB of NGC~6496 has become clearer.  However, a significant number
of field stars still contaminate this CMD, as evidenced by the CMD
shown in panel (c).  This figure includes only stars outside the
published tidal radius, so should display mainly field stars.  A CMD
of stars derived from the Galacitic population synthesis model of
\citet{robin03}\footnote{Obtained via
\tt{http://model.obs-besancon.fr}.} closely matches the observed CMD
in panel (c), supporting our claim that panel (c) is a good
representation of the field stars toward NGC~6496.

\begin{figure}
\epsscale{1.0}
\plotone{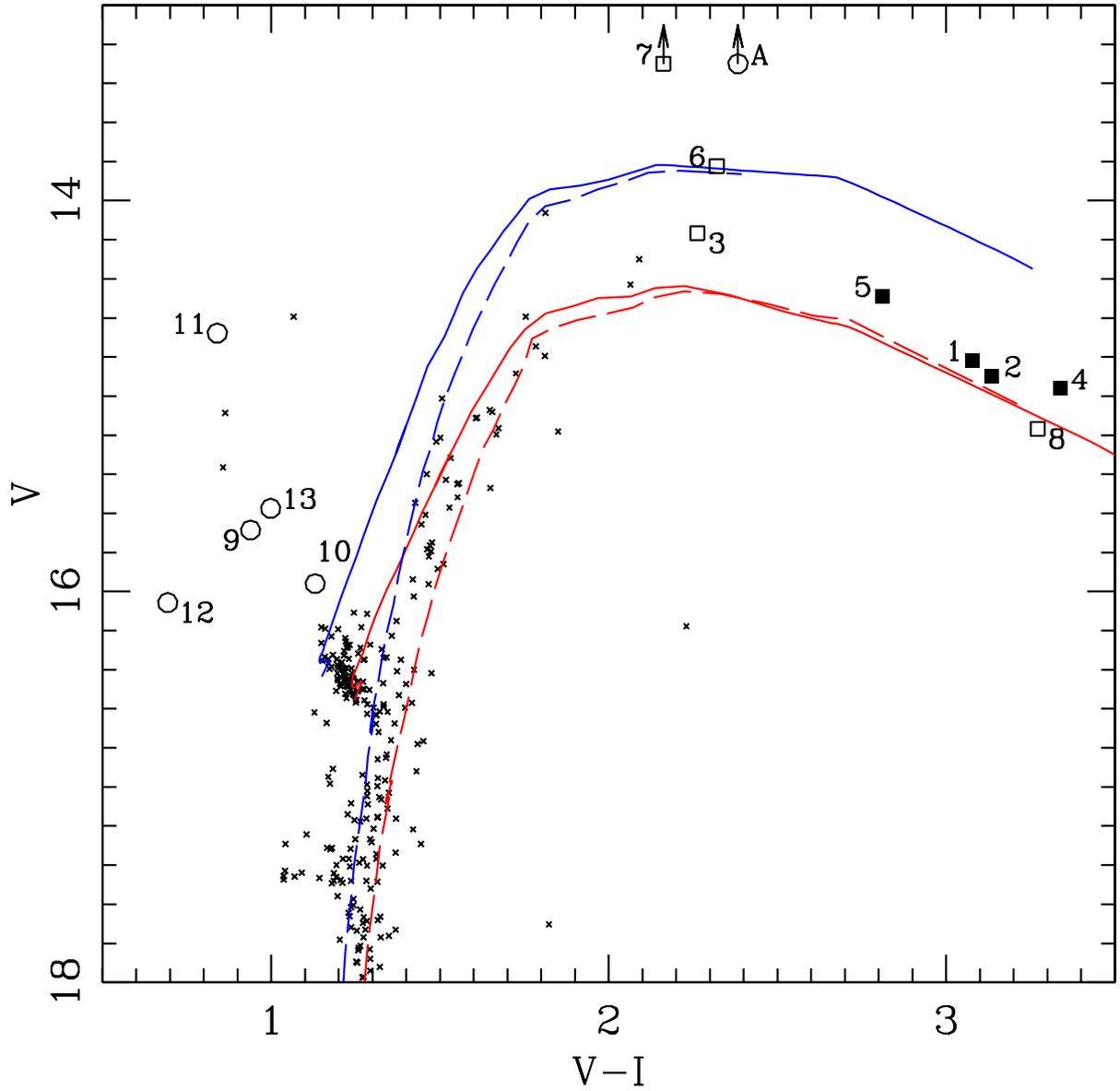}
%\plotone{fig_cmdvar.eps}
\caption{The field-subtracted color-magnitude diagram of stars within
2.5 arcmin of the center of NGC~6496, as described in
Sec. \ref{sec_cmd}.  Isochrones from \citep{girardi02} having ages of
11.2 Gyr and [Fe/H] = --0.7 (upper) and --0.4 (lower) are shown, with
the RGB dashed and the AGB solid.  Short-period variables are indicated
with circles and LPVs with squares; label numbers give the variable star 
number, ``V\#''.  The bright variables V7 ($V = 12.85$ mag)
and ASAS 175901-4411.5 (labeled "A", having $V = 11.31$ mag) are plotted
at fainter magnitudes for convenience in scaling the plot.
%Variable stars are indicated with
%the following symbols: LPVs (squares), short-period variables
%(circles), and variables brighter than $V = 13$ mag (the vertical limit of the plot (triangles).  
Solid and open symbols are used for variables located
inside and outside a projected radius of 2.5 arcmin from the cluster
center, respectively; the former are more likely to be cluster members.  
\label{fig_cmdvar}}
\end{figure}

Our clearest picture of the cluster CMD comes in Figure
\ref{fig_cmdvar}, where we have statistically subtracted field stars
from the CMD shown in Figure \ref{fig_cmd3}b.  Specifically, we
selected stars located in an annulus between 5.0 and 5.6 arcmin from
the cluster center to represent a sample of field stars.  The area of
this annulus is equal to the circular area of 19.6 arcmin$^2$ from
which stars in panel (b) were drawn.  For each star in the field
sample, we located the nearest star in color-magnitude space from panel
(b), and removed it.  The stars that remain should represent,
statistically, only stars belonging to the cluster.  Indeed, the stars
seen in Figure \ref{fig_cmdvar} are consistent with a metal-rich
globular cluster with only a few un-subtracted field stars falling off
the principal sequences.

We include in Figure \ref{fig_cmdvar} isochrones from
\citet{girardi02} for an age of 11.2 Gyr and metallicities bracketing
that of NGC~6496, [Fe/H] = --0.7 and --0.4 dex.  The cluster red giant
stars fall closer to the latter isochrone, as expected for a cluster
of metallicity of [Fe/H] = $-0.46$ dex \citet{harris96}.  Initial
isochrone placement using the reddening and apparent distance modulus
values from \citet{harris96} did not match well with the location of
the HB.  We found that $E(B-V) = 0.21 \pm 0.02$ and $(m-M)_V = 15.60
\pm 0.05$ mag produced a better match to the HB, assuming the value of
[Fe/H] = $-0.46$ dex is correct.  This fit is shown in Figure
\ref{fig_cmdvar}.  The errors indicate the range of horizontal and
vertical shifts that produce an acceptable visual fit to center of the
differentially reddened HB.  An additional contribution of $\sim$0.15
mag should be included in the distance uncertainty to reflect the 
luminosity calibration of the isochrones \citep{girardi02,dotter08}.

Our mean reddening value is in excellent agreement with that of
\citet{saraj94}, $E(B-V) = 0.22$ mag, who used a simultaneous solution
for reddening and metallicity to the shape and position of the bright
sequences in the CMD, and with the recalibrated dust maps of
\citet{schlafly11}, $E(B-V) = 0.20$ mag.  Reddening estimates using
integrated cluster light, such as the value $E(B-V) = 0.09$ mag from
\citet{zinnwest84}, may have been biased by the light of bright, blue
field stars seen in Figure \ref{fig_cmd3}b, which contribute strongly
to the light near 3900 \AA.  Such values in turn shifted Harris's mean
value of $E(B-V) = 0.15$ mag, complied from several literature
sources, away from what we believe to be the correct reddening.  Our
apparent distance modulus is significantly shorter than that in
\citet{harris96}, $(m-M)_V = 15.74$ mag.  This probably results from
the different photometric calibration we employ, as discussed in
Sec. \ref{sec_daophot}.
%
%The distance to NGC~6496 is 9.7 kpc under these assumptions,
%compared to 11.3 kpc from \citep{harris96}, though much of the
%difference is due to the different luminosity calibrations of the HB
%used by \citet{girardi02} and \citet{harris96}. (SKIP d?)

%%%%%%%%%%%%%%%%

\section{Variable Stars}   \label{sec_varstar}

We used the instrumental, time-series photometry obtained in separate
filters to distinguish variable from non-variable
sources. Specifically, we used the variability index, $\Lambda$,
calculated by DAOMASTER \citep{stetson94}, and plotted $\Lambda$
versus magnitude for each of the five data sets ($V$ and $I$ from
CTIO, and $V$, $I_L$, and $I_S$ from P4) to visually rank the
likelihood of variability of each star.  The magnitude of $\Lambda$
tended to correlate across the five data sets.  We extracted the
instrumental photometry of 18 stars having the largest $\Lambda$
rankings, and plotted their magnitude versus time plots to visually
separate stars with coherent variations from those with predominantly
constant magnitudes and occasional anomalous values, which usually
result from photometry compromised by crowding or an undetected image
defect.  Given the size of their photometric errors, stars with $V <
15$ mag (LPVs) could be detected with confidence for $V$ amplitudes as
low as $\sim$0.1 mag, while the detection amplitude increases
gradually for fainter variables.  Using this approach, we detected
eleven variable stars in both the CTIO and the P4 data sets, plus two
additional variables in the CTIO data that were located outside the P4
field of view.

Table \ref{tab_coord} shows positional information for these thirteen
variable stars, which we name V1 through V13.  Columns 2-4 show the
identification number and the ($x$, $y$) coordinates of each variable
star from our P4 photometry from Table \ref{tab_cmd}.
%\footnote{A FITS-format image corresponding to
%these coordinates is available as a finder chart, along with other
%data products from this study, at {\tt
%http://physics.bgsu.edu/$\sim$layden/publ.htm }.} 
The projected angular distance from the cluster center in arcminutes,
$R_{proj}$, is shown in the fifth column.  We matched the positions of
these variable stars on our images with images from the Two Micron
All-Sky Survey (2MASS, \citet{strutskie06}) to find rough coordinates,
then extracted precise equatorial coordinates and magnitudes in the
$J$ and $K$ passbands (and their uncertainties, $\sigma_J$ and
$\sigma_K$) from the 2MASS Point Source Catalog.  These values are
listed on the right side of Table \ref{tab_coord}.  All thirteen stars
were relatively uncrowded, and their ``phot\_qual'' flags in the 2MASS
data base were ``AAA,'' indicating the highest quality infrared
photometry.

We used differential photometry to calibrate our instrumental
time-series magnitudes of these variables.  For each variable, we
selected up to ten nearby, non-variable stars with colors and
magnitudes as close as possible to that of the variable to serve as
comparison stars
%\footnote{By doing differential photometry between a variable and
%nearby comparison stars, we reduce the effect of the
%spatially-dependent, quadratically-varying photomtric offsets in the
%CTIO data described in Sec. \ref{sec_daophot}.}.  
For each comparison star, we calculated an estimate of the variable
star's magnitude via
\begin{equation}  \label{coeff_vv}
V_v = (v_v - v_c) + V_c - \beta_1\left[ (V-I)_v - (V-I)_c \right],
\end{equation}
and an analogous equation for $I_v$, where the $v$ and $c$ subscripts
refer to the variable and comparison star, and $\beta_1$ is the color
term from Eqn. \ref{coeff_v}.

This results in $\sim$10 different standard magnitude estimates for
each variable star on each image.  We adopted the median of these
estimates as the final magnitude associated with that image, and the
standard error of the mean as the final uncertainty.  The use of
differential photometry with comparison stars located near the
variable helps to reduce any spatial-dependence of the photometry, as
discussed in Sec. \ref{sec_daophot}.

Table \ref{tab_timeser} presents the calibrated time-series photometry
of the variable stars we have detected.  The columns are, (1) variable
star designation, (2) observation time expressed as heliocentric
Julian Date minus 2,450,000.0 days, (3) the light curve phase for
stars with cyclic behavior, (4-6) the magnitude, error, and number of
comparison stars used for the $I_L$ photometry, (7-9) the magnitude,
error, and number of comparison stars used for the $I_S$ photometry,
and (10-12) the magnitude, error, and number of comparison stars used
for the $V$-band photometry.
%For each filter, the magnitude is the median
%of the $N$ comparison stars used, and the error is the standard error
%of the mean of those $N$ magnitudes (REWORD once discussion of
%differential photometry is in place).
% which err should I show??  NB: err1=mean of typical errors, err2=sem
% of N diffl photom values; we used err2 for LCs and in online table (tab2).

% Moe, despite your careful work on ISIS, I don't think it adds much
% to this paper (it was a good learning experience, though!), since
% the stars are so un-crowded.  We prefer DAOPHOT because it gives
% calibrated magnitudes rather than flux differences.

Image subtraction was also performed using the ISIS2.2 software
package \citep{alard00}.  The results confirmed the locations of the
DAOPHOT variables, but resulted in no additional variable star
candidates, so we do not discuss it further.

The positions of the variables in the CMD are shown in Figure
\ref{fig_cmdvar}.  Stars V1--V8 have locations consistent with being
long period variables, and the time-magnitude plots of these stars
further support their being LPV stars.  Similarly, the colors and
time-magnitude plots of V9-V13 indicate they have much shorter
periods, typical of RR Lyrae, type II Cepheid, or binary stars.

\subsection{Long Period Variables} \label{ssec_lpv}

The light curves of the long period variable stars V1--V6 are shown in
Figure \ref{fig_6lpv}.  The CTIO observations are shown on the left of
each panel, and the two years of P4 observations on the right with a
short gap around JD = 2,455,200 days when the cluster was below the
horizon all night.  All six LPVs have relatively low amplitudes and
irregular light variations.  Variables V7 and V8 were off the P4 field
of view, so we have only a short span of data for each star; enough to
identify them as LPVs, but too little to characterize them.

Table \ref{tab_varmags} presents the mean magnitudes ($\langle V
\rangle$ and $\langle I_L \rangle$) and light extrema ($V_{max}$,
$V_{min}$, $I_{max}$, $I_{min}$) for the LPVs, along with the number
of $V$-band images obtained, the derived periods and their
uncertainties, and the variability type.  We used several methods to
estimate pulsation periods, including phase dispersion minimization
(PDM; \citealt{stellingwerf78}), a template fitting method
%\citealt{layden1998,layden99,layden00}), and for the irregular
\citep{layden99}, and for the irregular pulsators, simply
averaging the time intervals between peaks.  For V1-6, the periods
listed are the mean from the three different methods, and the
uncertainty is their standard deviation.

The LPV V1 has a rather regular periodicity around 69 days, though the
amplitude waxes and wanes from one cycle to the next, with a maximum
range of $\Delta V = 1.3$ mag.  This behavior suggests the star is a
semi-regular (SR) pulsator.  Its proximity to the cluster center and
position on the CMD suggest it is a member of NGC~6496.

The LPVs V2-V6 show less stable periodicity, and amplitudes that are
smaller and quite variable, suggesting they are irregular (Lb)
pulsators.
% The distinction between SR and Lb may be more of degree along
% a continuum of pulsation properties than ...
% Lebzelter ? and Percy ?
The projected radii in Table \ref{tab_coord} and their locations along
the expected RGB/AGB of NGC~6496 in the CMD suggest that V2-5 are also
members of the cluster.  While V6 is within the cluster's tidal radius
of 4.8 arcmin, it is $\sim$0.5 mag brighter that the cluster sequence
and may be a field star.  Almost certainly, V7 and V8 are field stars,
as they are at or outside 4.8 arcmin and off the cluster sequence.
Obtaining spectra for radial velocities and metallicities would
clarify these membership estimates.

The LPV V3 shows a short-period, low-amplitude variation on top of a
much longer periodicity.  We estimate the long period to be $340 \pm
10$ days based on the one full cycle in our data, and after removing
that trend, found the short period to be rather stable at $37 \pm 1$
days.  The range of the slow pulsation is about 0.5 mag in $V$, while
that of the short period is about 0.2 mag.  Comparing this behavior
with those of stars discussed by \citet{percy04} and \citet{wood04},
we interpret the short period to represent the radial pulsation of the
star, while the longer period corresponds to a ``long secondary
period.''  \citet{wood04} found these long periods difficult to
explain from a theoretical view, though the most likely cause is a
combination of low-order non-radial modes and star spot activity.  As
our survey of globular clusters continues, we will compare variables
with long secondary periods in hopes that controlling key
observational variables will further constrain the cause of the long
secondary periods.

We can compare the LPVs in NGC~6496 to those in well-studied systems
of similar metallicity.  For example, \citet{wood99} showed that LPVs
grouped into distinct sequences in the period-luminosity diagram for a
large sample of LPVs in the Large Magellanic Cloud (LMC).  Our Figure
\ref{fig_logp} is the equivalent plot for the LPVs in NGC~6496, where
the luminosity is expressed as Wood's nearly reddening-free parameter
\begin{equation}  \label{coeff_iw}
I_W = \langle I \rangle - A_I - 1.38\left[\langle V \rangle - \langle I\rangle  - E(V-I) \right]
\end{equation}
in the top panel, and as the $K_s$-band absolute magnitude in the
bottom panel.  For the latter, we converted the single-epoch 2MASS
data from Table \ref{tab_coord} using the reddening and distance
modulus for the cluster.  Following \citet{lebwood05}, we estimated
the expected $K$-band magnitude range of each variable star as 20\% of
its $V$-band magnitude range from Table \ref{tab_varmags}.  This
serves as an estimate of the vertical uncertainty of the single-epoch
magnitude in this diagram; fortunately, most of these uncertainties
are small, as shown by the errorbars in Figure \ref{fig_logp}.
 
We also plot boxes representing the LMC-star sequences of
\citet{wood99} (top) and \citet{ita04} (bottom), which have been
shifted for the reddening and distance modulus of the LMC.
\citet{wood99} identified sequences C and B with Mira and semi-regular
LPVs, respectively; sequence D contained stars exhibiting a long
secondary period, while sequence E contained red giant contact
binaries and higer mass post-AGB stars; the physical cause of stars in
sequence A was not apparent.  In the upper panel, V1-6 fall in or near
Sequence B, and roughly match the model for LMC stars pulsating in the
first overtone, $P_1$ \citep{wood99}.  The semi-regular LPV V1 (the
central of the four solid squares) lies in the middle of Sequence B,
on the $P_1$ model, while the irregular LPVs surround it (their period
uncertainties may cause them to scatter away from their true positions
in this diagram).  The long secondary period of the faintest LPV, V3,
falls nicely in Sequence D with its LMC counterparts.

In the lower panel, the LPVs in NGC~6496 are compared with the boxes
defined by \citet{ita04}, who resolved Wood's A and B sequences into
five separate groups labeled A$\pm$, B$\pm$, and C$^\prime$.  The
division between the + and -- sub-sequences occurs at the RGB tip. The
C$^\prime$ sequence was thought to contain high-amplitude Mira
variables pulsating in the first overtone, distinct from the
semi-regular stars in the B$\pm$ sequence.  The F sequence contains
Cepheids pulsating in the fundamental mode.  Again, there is rough
agreement between the LPVs in NGC~6496 and the LMC.  In both panels,
the cluster variables are restricted to lower luminosities, since the
LMC contains younger, more massive stars that populate the high
luminosity ends of each sequence (the LMC boxes in the bottom panel
extend upward to $M_K = -8.5$ to $-9.0$ mag).  \citet{wood99} and
\citet{ita04} identified carbon-rich Miras as those with $(J-K)_0 >
1.4$ mag; no such stars exist in NGC~6496 based on the 2MASS
photometry in Table \ref{tab_coord}.
%None of our stars appear to be carbon stars, since they all have 
%$(J-K)_0 < 1.4$ mag.  
Though the bulk of the LMC stars have a composition close to that of
NGC~6496, there may be star-to-star composition differences among the
LMC stars that complicate further comparison.

The LPVs in the Galactic globular cluster 47 Tucanae from
\citet{lebwood05} provide another point of comparison.  These stars,
shown as crosses in Fig. \ref{fig_logp}, have a similar age and AGB
mass to those in NGC~6496, but have a slightly lower metallicity,
[Fe/H] = --0.72 \citep{harris96}.  In both panels, our stars are, on
average, more luminous than the 47 Tuc stars at a given period,
suggesting that metallicity is positively correlated with luminosity.
A similar correlation was found by \citet{fw00} among Mira variables
in globular clusters.  The slope of their relation, ${\Delta\log P} /
{\Delta}$[Fe/H] = 0.28 predicts an offset of ${\Delta\log P} = 0.07$
between NGC~6496 and 47~Tuc, consistent with the shifts seen in Figure
\ref{fig_logp}, suggesting the relation extends to the less-evolved
semi-regular variables farther from the AGB tip.

% \citet{fw00] Feast, M. \& Whitelock, P. 2000, in "The Evolution of the Milky Way: stars versus clusters," edited by F. Matteucci \& F. Giovannelli (Kluwer: Dordrecht), 229

As in 47 Tuc, the LPVs in NGC~6496 are concentrated in Sequences
B$\pm$ (semi-regular LPVs) and C$^\prime$ (first overtone Miras), with
no stars in Sequence A$\pm$.  While 47 Tuc has five Mira variables on
Sequence C (and possibly two stars evolving into this fundamental-mode
sequence from the low-amplitude, lower-luminosity LPVs in Sequences
B$\pm$ and C$^\prime$), NGC~6496 has none.  However, 47 Tuc is richer
in LPV stars, and the ratio of 47 Tuc stars in these two regimes is
4-5 to one, so it is not significant that NGC~6496 currently harbors
no Mira variables.  None of the 47 Tuc LPVs shown in Figure
\ref{fig_logp} fall in Sequence D like V3 in NGC~6496, but about 30\%
of the 47 Tuc LPVs have a secondary ``long'' period noted, but not
quantified, in Table 1 of \citet{lebwood05}; a percentage consistent
with that seen in NGC~6496.

The curves shown in the lower panel are models from \citet{lebwood05}
showing pulsating AGB stars subjected to mass loss (see their
Fig. 6b).  Most of the stars in both NGC~6496 and 47~Tuc seem to lie
on the model line for second-overtone pulsators, $P_2$.  The two model
comparisons shown in Fig. \ref{fig_logp} lead us
to think that the LPVs in NGC~6496 are pulsating in the first or
second overtone.

% Wood, P. R. et al.(1999)]{wood99} Wood, P. R., Alcock, C. et
% al. 1999, in IAU Symp. 191,  Asymptotic Giant Branc Stars, eds. Le
% Bertre, Lebre, \& Waelkens, 151

%Add comment about Matsunaga IR measurements of LPVs in other clusters?

\subsection{Short Period Variables}

As previously noted, the blue colors ($V-I < 1.2$) and
scattered time-magnitude plots of variables V9-V13 indicate they have
short periods.  We used PDM and the template fitting method to search
for the periods of these variables.  We found acceptable periods which
are listed in column 9 of Table \ref{tab_varmags}.  The uncertainty in
the periods, determined as described in the section above is listed in
column 10.  The larger uncertainties for V11 and V12 reflect the
smaller number of observations, which led to multiple minima of similar
depth in the period diagnostic statistics.
%This table also
%contains the intensity-mean magnitudes of each star, $\langle V
%\rangle$ and $\langle I \rangle$, the maximum and minimum brightness
%in $V$ and $I$, and the number paired $VI$ observations.  All of our short period variables have relatively blue
%colors ($V-I$ $\textless$ 1.2).
Figure \ref{fig_6spv} shows the resulting phased light curves for V9--V13.

Of these stars, the light curves of V10 and V13 are easiest to
classify.  The continuous light variations and equal depth minima of
V10 mark it as a W~Ursae Majoris (EW) type contact binary.  While the
light variations of V13 are also continuous, the differing depths of
the primary and secondary eclipses and the primary eclipse depth
($\Delta V > 1.0$ mag) indicate it is a Beta Lyrae (EB) ellipsoidal
eclipsing binary.  The light curve properties of V9 seem closer to
those of V10 than V13 (the best fitting template was of an EW star),
but the apparent asymmetries around the secondary minimum make a
definitive classification difficult; we provisionally classify V9 as a
W UMa eclipsing binary.  The sparse data on V11 and V12, and the
consequent uncertainties in the period, make the classification of
these stars even more difficult.  Based on their light curve shapes,
amplitudes, colors, and best-fit templates, we provisionally classify
them as W UMa eclipsing binaries as well.

The locations of these stars in the CMD in Figure \ref{fig_cmdvar}
suggest they are members of the foreground field population, rather
than cluster members.  Also, V10, V11, and V13 are outside the
cluster's tidal radius, while V9 is close to it.  We argue that all
five short period variables are field stars.

\subsection{Other Variables}

As noted in the Introduction, previous searches for variable stars in
NGC~6496 found no candidates \citep{fourcade66, clement01}.
Sometimes, field star searches for variable stars identify cluster
variables.  We therefore searched the International Variable Star
Index database\footnote{See \tt{http://www.aavso.org/vsx/}.} for
objects within 8 arcmin of the cluster center, but found only one
object,
%no known variable stars.  Searching the All Sky Automated Survey 
%(ASAS) database \citep{asas} within the same radius yielded one star, 
ASAS 175901-4411.5 (see Table \ref{tab_coord}), from the All Sky
Automated Survey \citep{asas}.
%, located at $\alpha$ = 17:59:01 and $\delta$ = -44:11:30 (J2000).  
The automatic light curve characterization performed by ASAS yielded a
mean $V$ magnitude of 11.14, a $V$ amplitude of 0.55 mag, and a period
of 392 days.  The scatter in the phased light curve was large, and the
variable type automatically assigned was ``MISC,'' indicating a
solution insufficient to classify the variable type.

However, our inspection of the phased light curve suggested the star
may be an eclipsing binary with about twice the ASAS period.  We used
PDM to find $P = 740 \pm 10$ days, as shown in Figure \ref{fig_6spv}.
Though the scatter is still large, there appear to be primary and
secondary eclipses with slightly different depths, $V_{min1} = 11.75$
and $V_{min2} = 11.60$, respectively.  We therefore think this star is
a near-contact eclipsing binary containing two red giant stars,
analogous to a W UMa (EW) eclipsing binary containing two dwarf stars.

This star is saturated on most of our images, so we did not detect it
as a variable star.  However, it was unsaturated in enough $I_s$ and
poor-seeing $V$ images to appear in our photometric catalog as ID\#1,
with $V = 11.31$ and $V-I = 2.38$ mag.  This is much brighter than the
cluster giant branch stars and LPVs shown in Figure \ref{fig_cmdvar},
and the star is at a projected angular distance of 4.4 arcmin from the
cluster center, so we are quite certain the star is not a member of
the cluster.  Thus, we did not assign it a ``V'' number.

\section{Summary}   \label{sec_summary}

We have obtained time-series images in the $V$ and $I$ bandpasses for
a 10 arcmin field around the metal-rich globular cluster NGC~6496
using the CTIO 0.9-m and PROMPT4 telescopes.  We performed photometry
using the DAOPHOT~II and ALLFRAME packages. Photometric calibration
was obtained from 33 on-field photometric standard stars from
\citet{stetson00}.  Comparison with three published photometric
studies \citep{armand88,richtler94,saraj94} suggests our mean $VI$
photometry for NGC~6496 is the best currently available (see Table
\ref{tab_cmd}).  Comparison with isochrones suggests the reddening and
apparent distance modulus of the cluster are $E(B-V) = 0.21 \pm 0.02$
and $(m - M)_V = 15.60 \pm 0.15$ mag, assuming [Fe/H] = --0.46 dex.

\begin{figure}
\epsscale{1.0}
\plotone{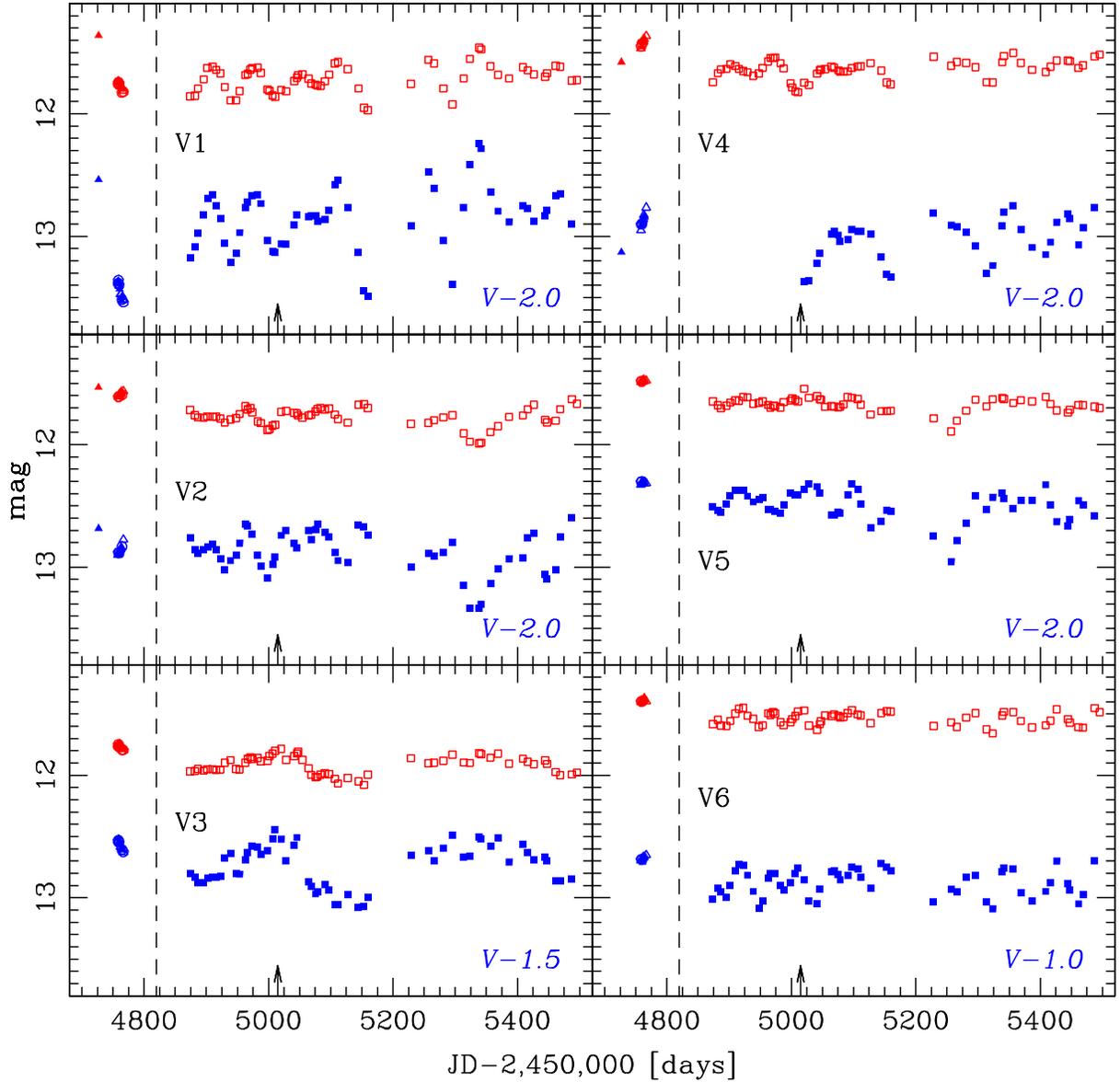}
%\plotone{lc_6lpv.eps}
\caption{Light curves for the long period variables V1--V6.  In each
panel, the upper light curve is $I_L$ and the lower curve is $V$,
shifted upward as indicated for convenience of display.  The data to
the left of the dashed line were taken at CTIO and are shifted by 4500
days in Julian Date for display.  The vertical arrows mark the date at
which the P4 image sets were separated due to camera rotation.
Photometric errors are smaller than the plot symbols.
\label{fig_6lpv}}
\end{figure}

Although previous searches for variable stars in NGC~6496 found no
candidates, we detected thirteen variable stars that are listed in
Table \ref{tab_coord}.  We provide periods for most of the stars,
along with light curve characteristics and variability types.
%While the  periods for these stars were
%calculated using PDM and the template fitting method, the
%classification was performed using the template fitting method in
%addition to the visual inspection of the light curves phased to
%different trial periods.
Six of the variable stars, V1--V6, are classified as LPVs (one
semi-regular and five irregular types) while five stars (V9--V13) are
short period variables (EW contact and EB ellipsoidal eclipsing
binaries).  Although we detected long-term variability in V7 and V8
consistent with their being LPVs, our data were insufficient to
determine their periodicity or light curve properties; we suggest
follow up observations for these stars. Searching the International
Variable Star Index for variable stars near our cluster yielded one
object (ASAS 175901-4411.5) with unknown type of variability. Our
color data and period analysis suggest this star is a contact
eclipsing binary containing two red giant stars with $P = 740 \pm 10$
days. The phased light curves of our variable stars are plotted in
Figure \ref{fig_6lpv} and Figure \ref{fig_6spv}.  Of the LPVs, V1-V5
appear to be members of NGC~6496, V6 may be a member, while V7, V8,
and all of the short period variables appear to be field stars.

\begin{figure}
\epsscale{1.0}
\plotone{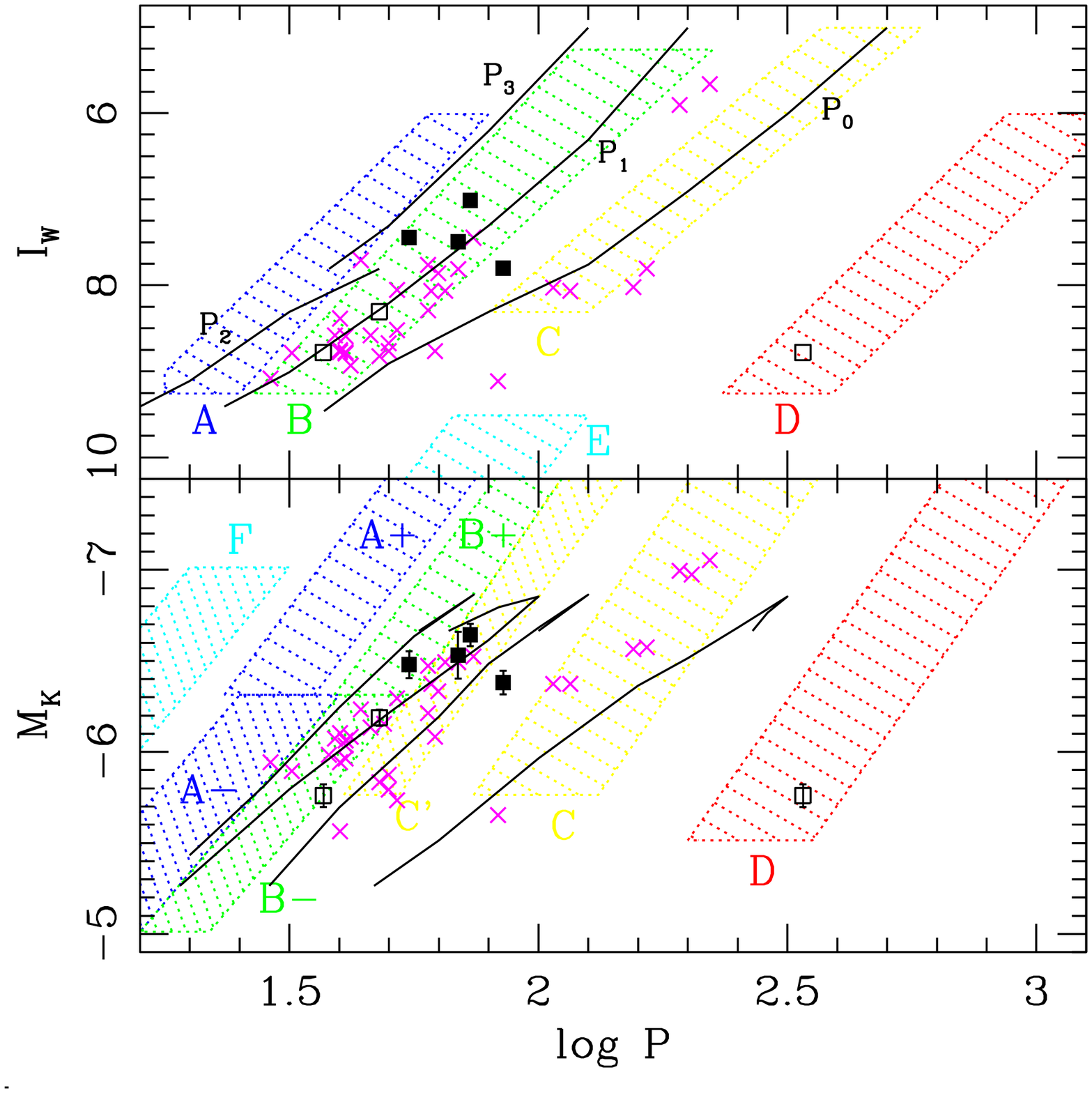}
%\plotone{fig_mag_logp.eps}
\caption{Period-luminosity plots showing (top) $I$-band and (bottom)
$K$-band magnitudes, as described in Sec. \ref{ssec_lpv}.  Squares
mark LPVs V1-V6 in NGC~6496 inside (filled) and outside (open) a projected
radius of 2.5 arcmin, while crosses mark the locations of LPVs in 47
Tuc \citep{lebwood05}.  The hatched regions mark the sequences of LPVs
in the Large Magellanic Cloud shown by \citet{wood99} (top) and
\citet{ita04} (bottom).  In both panels, the faintest LPV in NGC~6496,
V3, appears in both sequences B and D.  The curves in the top panel
show pulsation models for LMC stars from \citet{wood99} for the
fundamental mode ($P_0$) and the first through third overtones
($P_1-P_3$).  The curves in the lower panel are $P_0-P_3$ models (right to
left) for pulsating AGB stars undergoing mass loss \citep{lebwood05}.
\label{fig_logp}}
\end{figure}

\begin{figure}
\epsscale{1.0}
\plotone{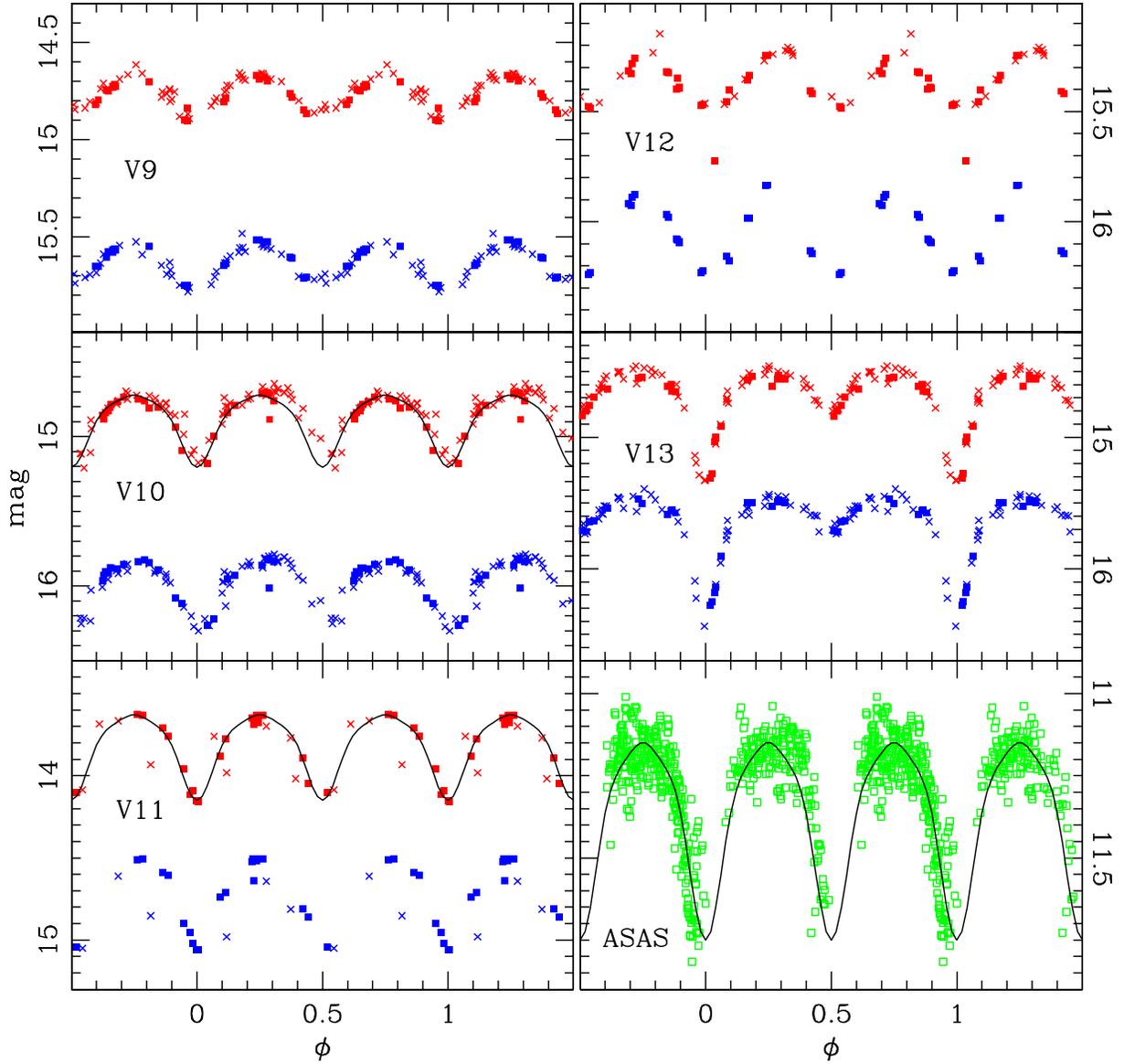}
%\plotone{lc_6spv.eps}
\caption{Phased light curves for the short period variables V9--V13
and ASAS 175901-4411.5, with two cycles shown for continuity.  In the
panels for V9--V13, the upper light curve is $I_L$ and the lower curve
is $V$, while the solid squares and crosses indicate CTIO and P4 data,
respectively.  Photometric errors are smaller than the plot symbols.
For ASAS 175901-4411.5, the data from the ASAS project are shown.
Several panels show the best-fitting template for a W~UMa contact
binary star.
%Error bars are included on points over one cycle.
%No period was found for V14 so magnitude vs. time is shown.  
\label{fig_6spv}}
\end{figure}

We compare the period-luminosity distribution of our LPVs to those of
LPVs in the LMC and 47 Tuc in Figure \ref{fig_logp}. 
%Our comparison shows that our V1-6 are located close to the model for 
%LPVs pulsating in the first overtone in the LMC. 
Our LPVs are located in or near the LMC-star Sequence B where stars
are thought pulsate in the first overtone mode with low amplitudes.
Our period-luminosity distribution of LPVs also mimic the distribution
of LPVs in 47 Tuc, though our stars appear slightly brighter,
suggesting a positive correlation between luminosity and metallicity
for these LPVs.  Most of the stars in NGC~6496 and 47 Tuc are located
close to the models for stars pulsating in either the first or second
overtone, depending on whether models exclude or include AGB
mass-loss, respectively.  The absence of Mira variables in NGC~6496
can be explained by the cluster's small stellar content relative to 47
Tuc.

%The main difference
%between LPVs period-luminosity distribution in NGC~6496 and 47 Tuc are
%the five Mira stars found in 47 Tuc and located in Sequence C
%(pulsations in the fundamental mode) in compared to no Mira stars
%found in NGC~6496. Because 47 Tuc has more stars than NGC~6496 and
%because the ratio of non-Mira LPVs to Mira stars in 47 Tuc is 4-5, it
%is not surprising that none of our six LPVs is a Mira star. 

We recommend spectroscopic observations of the LPVs in NGC~6496 to
further constrain their membership.  Because obtaining a complete
census of LPVs, along with accurate periods, light curve properties,
and evolutionary states helps in understanding and improving
theoretical pulsation models, stellar interior properties, and AGB
mass loss, we are continuing long-term photometric monitoring of about
twenty Galactic globular clusters that span a range of metallicities.


%% Included in this acknowledgments section are examples of the
%% AASTeX hypertext markup commands. Use \url without the optional [HREF]
%% argument when you want to print the url directly in the text. Otherwise,
%% use either \url or \anchor, with the HREF as the first argument and the
%% text to be printed in the second.

\acknowledgments

We appreciate the helpful recommendations of an anonymous referee.
Observations using PROMPT were made possible by the Robert Martin
Ayers Science Fund.  M.A. acknowledges support by the Collaborative
Research Center ``The Milky Way System'' (SFB 881, subproject A3) of
the German Research Foundation (DFG).  A.L. acknowledges support from
the U.S.  National Science Foundation under Grant No. 9988259 and from
NASA through Hubble Fellowship grant HF-01082.01-96A, which was
awarded by the Space Telescope Science Institute.  D.W. acknowledges
support from the Natural Sciences and Engineering Research Council of
Canada (NSERC) in the form of a Discovery Grant.  This publication
makes use of data products from the Two Micron All Sky Survey, which
is a joint project of the University of Massachusetts and the Infrared
Processing and Analysis Center/California Institute of Technology,
funded by the National Aeronautics and Space Administration and the
National Science Foundation.  This research has made use of the
International Variable Star Index (VSX) database, operated at AAVSO,
Cambridge, Massachusetts, USA.

\clearpage

%% the file, uncomment the \plotone, \plottwo, and
%% If you need a layout that cannot be achieved with \plotone or
%% \plottwo, you can invoke the graphicx package directly with the
%% \includegraphics command or use \plotfiddle. For more information,
%% please see the tutorial on "Using Electronic Art with AASTeX" in the
%% documentation section at the AASTeX Web site,
%% http://www.journals.uchicago.edu/AAS/AASTeX.
%%
%% The examples below also include sample markup for submission of
%% supplemental electronic materials. As always, be sure to check
%% the instructions to authors for the journal you are submitting to
%% for specific submissions guidelines as they vary from

%% This example uses \plotone to include an EPS file scaled to
%% 80% of its natural size with \epsscale. Its caption
%% has been written to indicate that additional figure parts will be
%% available in the electronic journal.

\clearpage

\clearpage

\clearpage

\clearpage

\clearpage

\clearpage

\clearpage

\clearpage

%% Here we use \plottwo to present two versions of the same figure,
%% one in black and white for print the other in RGB color
%% for online presentation. Note that the caption indicates
%% that a color version of the figure will be available online.
%%

%% If you are not including electronic art with your submission, you may
%% mark up your captions using the \figcaption command. See the
%% No more than seven \figcaption commands are allowed per page,
%% so if you have more than seven captions, insert a \clearpage

%% Tables should be submitted one per page, so put a \clearpage before

%% Two options are available to the author for producing tables:  the
%% deluxetable environment provided by the AASTeX package or the LaTeX
%% table environment.  Use of deluxetable is preferred.
%%

%% Three table samples follow, two marked up in the deluxetable environment,
%% one marked up as a LaTeX table.

%% In this first example, note that the \tabletypesize{}
%% command has been used to reduce the font size of the table.
%% We also use the \rotate command to rotate the table to
%% landscape orientation since it is very wide even at the
%% Note also that the \label command needs to be placed

%% This table also includes a table comment indicating that the full
%% version will be available in machine-readable format in the electronic

%\begin{deluxetable}{ccrrrrrr}\    
% The trailing \ forces the table to the bottom of the page...

\begin{deluxetable}{ccrrrrrr}
\tabletypesize{\scriptsize}
%\rotate
\tablecaption{Transformation Coefficients\label{coeff}}
\tablewidth{0pt}
\tablehead{
Set & Filter & $\alpha$ & $\epsilon_{\alpha}$ & $\beta$ & $\epsilon_{\beta}$ & RMS }
\startdata
CTIO & $V$   &  0.273 & 0.004 &  0.011 & 0.010 & 0.033 \\
CTIO & $I$   &  0.759 & 0.007 &  0.034 & 0.016 & 0.052 \\
P4   & $V$   & -0.612 & 0.003 &  0.000  & 0.000  & 0.015 \\
P4   & $I_L$ &  0.310 & 0.005 & -0.075 & 0.007 & 0.020 \\
P4   & $I_S$ &  1.910 & 0.005 & -0.095 & 0.008 & 0.022 \\
\enddata
\end{deluxetable}

% see /physics1/people/layden/N6496/MOE_ABBAS/fitrot/sm.stetson_comp2014

\clearpage

\begin{deluxetable}{rrrrrrrrr}
\tabletypesize{\scriptsize}
%\rotate
\tablecaption{Mean Photometry of NGC~6496\label{tab_cmd}}
\tablewidth{0pt}
\tablehead{
ID & $X_{pix}$ & $Y_{pix}$ & $V$ & $\epsilon_{V}$ & $I$ &
$\epsilon_{I}$ 
%& L/S\tablenotemark{a} 
& Chi & Sharp }
\startdata
% 1 & 445.18 &  21.88 & 11.306 & 0.0056 &  8.923 & 0.0054 & 2 & 1.031 & -0.010 \\
% 2 & 714.37 &1022.18 & 10.275 & 0.0123 & 10.134 & 0.0097 & 1 & 1.709 &  0.035 \\
% 3 & 644.07 & -67.01 & 11.215 & 0.0073 &  9.814 & 0.0071 & 1 & 0.994 & -0.012 \\
% 4 & 491.00 & 439.53 & 11.379 & 0.0054 & 10.129 & 0.0050 & 1 & 1.574 & -0.049 \\
% 5 & -49.51 & 406.18 & 13.098 & 0.0078 &  9.591 & 0.0075 & 1 & 1.516 &  0.007 \\
% 6 & -34.54 & 639.31 & 12.907 & 0.0069 & 10.762 & 0.0065 & 1 & 1.328 & -0.006 \\
% 7 & 815.93 & 647.35 & 12.749 & 0.0054 & 11.304 & 0.0051 & 1 & 1.262 & -0.001 \\
 1 & 445.18 &  21.88 & 11.306 & 0.0056 &  8.904 & 0.0064 & 1.031 & -0.010 \\
 2 & 714.37 &1022.18 & 10.275 & 0.0123 & 10.096 & 0.0089 & 1.709 &  0.035 \\
 3 & 644.07 & -67.01 & 11.215 & 0.0073 &  9.812 & 0.0076 & 0.994 & -0.012 \\
 4 & 491.00 & 439.53 & 11.379 & 0.0054 & 10.122 & 0.0055 & 1.574 & -0.049 \\
 5 & -49.51 & 406.18 & 13.098 & 0.0078 &  9.619 & 0.0090 & 1.516 &  0.007 \\
 6 & -34.54 & 639.31 & 12.907 & 0.0069 & 10.757 & 0.0076 & 1.328 & -0.006 \\
 7 & 815.93 & 647.35 & 12.749 & 0.0054 & 11.301 & 0.0056 & 1.262 & -0.001 \\
\enddata
\tablecomments{Table \ref{tab_cmd} is published in its entirety in
the electronic edition of the {\it AJ}.  A portion is shown here for
guidance regarding its form and content.}
%\tablenotetext{a}{This flag indicates whether $I_L$ (1) or $I_S$ (2) data was used.}
\end{deluxetable}

\clearpage

\begin{deluxetable}{lrrrrllrrrr}
\tabletypesize{\scriptsize}
%\rotate
\tablecaption{Variable Star Coordinates and 2MASS Photometry\label{tab_coord}}
\tablewidth{0pt}
\tablehead{
\colhead{V\#} & \colhead{ID$_{P4}$} & \colhead{$X_{P4}$} & \colhead{$Y_{P4}$} &
\colhead{$R_{proj}$} &
\colhead{RA(J2000)} & \colhead{Dec(J2000)} & \colhead{$J$} & \colhead{$\sigma_J$} &
\colhead{$K$} & \colhead{$\sigma_K$}
% & \colhead{PQ\tablenotemark{b}}
}
\startdata
V1  &  31 & 541.4 & 300.4 & 2.08 & 17:58:55.88 & --44:14:19.4 &  9.810 & 0.024 &  8.492 & 0.023 \\
V2  &  32 & 240.7 & 619.3 & 2.21 & 17:59:12.94 & --44:17:20.5 &  9.890 & 0.024 &  8.543 & 0.019 \\
V3  &  29 & 145.0 & 527.3 & 2.64 & 17:59:17.99 & --44:16:24.4 & 10.458 & 0.022 &  9.263 & 0.019 \\
V4  &  30 & 528.0 & 458.1 & 1.14 & 17:58:56.91 & --44:15:51.7 &  9.661 & 0.024 &  8.381 & 0.027 \\
V5  &  27 & 419.0 & 285.8 & 1.79 & 17:59:02.55 & --44:14:08.3 &  9.944 & 0.024 &  8.642 & 0.021 \\
V6  &  18 & 549.3 & 839.8 & 3.85 & 17:58:56.47 & --44:19:36.4 & 10.035 & 0.024 &  8.835 & 0.021 \\
V7\tablenotemark{a}  & ... &  ...  &  ...  & 4.71 & 17:59:28.01 & --44:17:26.7 &  9.311 & 0.023 &  8.195 & 0.027 \\
V8\tablenotemark{a}  & ... &  ...  &  ...  & 5.76 & 17:59:05.82 & --44:21:39.1 & 10.064 & 0.022 &  8.807 & 0.019 \\
V9  & 256 & 930.6 & 334.7 & 4.21 & 17:58:34.68 & --44:14:47.1 & 14.274 & 0.029 & 13.893 & 0.043 \\
V10 & 244 &  61.9 & 841.3 & 5.89 & 17:59:23.14 & --44:19:27.2 & 14.028 & 0.036 & 13.371 & 0.035 \\
V11 & 115 &1009.2 & 543.8 & 4.95 & 17:58:30.75 & --44:16:51.4 & 13.265 & 0.024 & 12.776 & 0.023 \\
V12 & 418 & 358.1 & 168.2 & 2.60 & 17:59:05.67 & --44:12:58.0 & 14.885 & 0.040 & 14.623 & 0.073 \\
V13 & 226 & 841.0 & 808.8 & 4.81 & 17:58:40.44 & --44:19:23.9 & 13.952 & 0.054 & 13.419 & 0.053 \\
\enddata
%% Text for table notes should follow after the \enddata but before
%% the \end{deluxetable}. Make sure there is at least one \tablenotemark
%% in the table for each \tablenotetext.
%\tablecomments{The equatorial coordinates with high precision are the
%centroid positions from the 2MASS Point Source Catalog \citep{strutskie06}, while the
%lower precision coordinates were estimated from the 2MASS images using
%the world coordinate system solutions.}
\tablenotetext{a}{ Stars located outside the P4 field of view.}
\end{deluxetable}

\clearpage

\begin{deluxetable}{lrrrrrrrrrrr}
\tabletypesize{\scriptsize}
%\rotate
\tablecaption{Time Series Photometry\label{tab_timeser}}
\tablewidth{0pt}
\tablehead{
\colhead{V\#} & \colhead{Time\tablenotemark{a}} & 
\colhead{Phase}\tablenotemark{b} & 
\colhead{$I_L$} & \colhead{$\sigma_L$} & \colhead{$N_L$} & 
\colhead{$I_S$} & \colhead{$\sigma_S$} & \colhead{$N_S$} & 
\colhead{$V$} & \colhead{$\sigma_V$} & \colhead{$N_V$} 
% & \colhead{PQ\tablenotemark{b}}
}
\startdata
V1 &  226.7420 & 9.999 & 11.362 & 0.007 & 5 & 99.999 & 9.999 & 0 &
14.537 & 0.011 & 5 \\
V1 &  258.6157 & 9.999 & 11.737 & 0.002 & 5 & 99.999 & 9.999 & 0 &
15.376 & 0.003 & 5 \\
V1 &  258.6231 & 9.999 & 11.758 & 0.001 & 5 & 99.999 & 9.999 & 0 &
15.381 & 0.003 & 5 \\
... &  ...  &  ...  & ... &  ...  &  ...  & ... &  ...  &  ...  & ... &  ...  &  ...  \\
V1 & 5487.5210 & 9.999 & 11.729 & 0.006 & 9 & 11.748 & 0.009 & 9 & 14.896 & 0.007 & 9 \\
V1 & 5495.5190 & 9.999 & 11.725 & 0.006 & 9 & 11.749 & 0.007 & 9 & 99.999 & 9.999 & 0 \\
V2 &  226.7420 & 9.999 & 11.534 & 0.010 & 5 & 99.999 & 9.999 & 0 & 14.684 & 0.009 & 5 \\
V2 &  258.6157 & 9.999 & 11.606 & 0.002 & 5 & 99.999 & 9.999 & 0 & 14.898 & 0.003 & 5 \\
... &  ...  &  ...  & ... &  ...  &  ...  & ... &  ...  &  ...  & ... &  ...  &  ...  \\
V8 &  266.8369 & 9.999 & 11.882 & 0.003 & 5 & 99.999 & 9.999 & 0 & 15.092 & 0.003 & 5 \\
V9 &  226.7471 & 0.491 & 14.777 & 0.007 & 5 & 99.999 & 9.999 & 0 & 15.619 & 0.005 & 5 \\
V9 &  258.6157 & 0.060 & 14.899 & 0.005 & 5 & 99.999 & 9.999 & 0 & 15.749 & 0.004 & 5 \\
... &  ...  &  ...  & ... &  ...  &  ...  & ... &  ...  &  ...  & ... &  ...  &  ...  \\
V13 & 5487.5210 & 0.192 & 14.799 & 0.015 & 6 & 14.769 & 0.029 & 7 & 15.703 & 0.012 & 6 \\
V13 & 5495.5190 & 0.431 & 14.548 & 0.005 & 8 & 14.554 & 0.006 & 8 & 99.999 & 9.999 & 0 \\
\enddata
%% Text for table notes should follow after the \enddata but before
%% the \end{deluxetable}. Make sure there is at least one \tablenotemark
%% in the table for each \tablenotetext.
\tablecomments{Table \ref{tab_timeser} is published in its entirety in
the electronic edition of the {\it AJ}.  A portion is shown here for
guidance regarding its form and content.}
\tablenotetext{a}{Time of observation expressed as Julian Date - 2,450,000.0 days.}
\tablenotetext{b}{A value of 9.999 is used for LPV stars for which we
did not make period-phased light curves.}
\end{deluxetable}

\clearpage

\begin{deluxetable}{lcrrrrrrrrc}
\tabletypesize{\scriptsize}
%\rotate
\tablecaption{Variable Star Properties\label{tab_varmags}}
\tablewidth{0pt}
\tablehead{
\colhead{V\#}  & \colhead{$\langle V \rangle$} & 
\colhead{$\langle I \rangle$} & \colhead{$V_{max}$} & \colhead{$V_{min}$} & 
\colhead{$I_{max}$} & \colhead{$I_{min}$} & \colhead{$N_V$} & 
\colhead{$P$} & \colhead{$\sigma_P$}  & \colhead{Type}
% & \colhead{PQ\tablenotemark{b}}
}
\startdata
%V#    <V>    <I_L>    Vmax    Vmin    Imax    Imin   Nv   Per  Perr   type  
V1  & 14.98 & 11.72 & 14.25 & 15.54 & 11.36 & 11.97 & 72 &  69 & 2    & SR  \\
V2  & 14.87 & 11.74 & 14.60 & 15.34 & 11.53 & 11.99 & 70 &  55 & 3    & Lb  \\
V3  & 14.20 & 11.89 & 13.94 & 14.58 & 11.74 & 12.08 & 71 & 37\tablenotemark{a} & 1 & Lb  \\
V4  & 14.98 & 11.60 & 14.75 & 15.37 & 11.37 & 11.82 & 49 &  73 & 3    & Lb  \\  
V5  & 14.46 & 11.64 & 14.30 & 14.96 & 11.44 & 11.89 & 69 &  85 & 15   & Lb  \\
V6  & 13.84 & 11.51 & 13.65 & 14.09 & 11.36 & 11.66 & 69 &  48 & 1    & Lb  \\  
V7  & 12.85 & 10.69 & 12.82 & 12.88 & 10.67 & 10.70 &  9 & ... & ...  & Lb? \\
V8  & 15.16 & 11.90 & 15.08 & 15.22 & 11.88 & 11.93 & 16 & ... & ...  & Lb? \\
V9  & 15.63 & 14.76 & 15.51 & 15.78 & 14.61 & 14.90 & 61 & 0.76664 & 0.00005 & EW? \\
V10 & 15.96 & 14.83 & 15.79 & 16.30 & 14.65 & 15.21 & 78 & 0.28974 & 0.00001 & EW  \\
V11 & 14.76 & 13.85 & 14.50 & 15.06 & 13.63 & 14.16 & 27 & 0.3371  & 0.0020  & EW? \\
V12 & 16.04 & 15.35 & 15.83 & 16.25 & 15.21 & 15.48 & 34 & 0.884   & 0.001   & EW? \\
V13 & 15.64 & 14.68 & 15.43 & 16.44 & 14.46 & 15.33 & 78 & 0.43851 & 0.0002  & EB  \\
ASAS& 11.31 &  8.92 & 11.15 & 11.75 & ...   & ...   &490 & 740 & 10     & EW? \\
\enddata
%% Text for table notes should follow after the \enddata but before
%% the \end{deluxetable}. Make sure there is at least one \tablenotemark
%% in the table for each \tablenotetext.
%\tablecomments{Table \ref{tab_timeser} is published in its entirety in
%the electronic edition of the {\it AJ}.  A portion is shown here for
%%guidance regarding its form and content.}
\tablenotetext{a}{A long secondary period of $340 \pm 10$ d was found
for V3.}
\end{deluxetable}

%% If you use the table environment, please indicate horizontal rules using
%% Do not put multiple tabular environments within a single table.
%% The optional \label should appear inside the \caption command.

%% If the table is more than one page long, the width of the table can vary
%% from page to page when the default \tablewidth is used, as below.  The
%% individual table widths for each page will be written to the log file; a
%% maximum tablewidth for the table can be computed from these values.
%% The \tablewidth argument can then be reset and the file reprocessed, so
%% that the table is of uniform width throughout. Try getting the widths
%% from the log file and changing the \tablewidth parameter to see how
%% adjusting this value affects table formatting.

%% The \dataset{} macro has also been applied to a few of the objects to
%% show how many observations can be tagged in a table.

%\clearpage

%% Tables may also be prepared as separate files. See the accompanying
%% sample file table.tex for an example of an external table file.
%% To include an external file in your main document, use the \input
%% command. Uncomment the line below to include table.tex in this
%% sample file. (Note that you will need to comment out the \documentclass,
%% \begin{document}, and \end{document} commands from table.tex if you want
%% to include it in this document.)


%% The following command ends your manuscript. LaTeX will ignore any text


\begin{thebibliography}{}

\bibitem[Alard(2000)]{alard00} Alard, C. 2000, \aaps, 144, 363

\bibitem[Armandroff(1988)]{armand88} Armandroff, T. E. 1988, \aj, 96, 588

\bibitem[Baker et al. (2007)]{baker07} Baker, J. M., Layden, A. C.,
Welch, D. L., \& Webb, T. M. A. 2007, \aj, 133, 139

%\bibitem[Bailey(1924)]{bailey24} Bailey, S. I. 1924, Harvard Bull. 802, 2

%\bibitem[Castellani et al.(2003)]{ccc03} Castellani, M., Caputo, F., \&
%Castellani, V. 2003, \aap, 410, 871

\bibitem[Clement et al.(2001)]{clement01} Clement, C. M., Muzzin, A.,
Dufton, Q. et al. 2001, \aj, 122, 2587

\bibitem[Dotter et al.(2008)]{dotter08} Dotter, A., Chaboyer, B.,
Jevremovi\'{c}, D., Kostov, V., Baron, V., \& Ferguson, J.W. 2008,
\apjs, 178, 89

%\bibitem[Feast(1973)]{feast73} Feast, M. W. 1973, IAU Colloq. 21,
%edited by J. D. Fernie, (Reidel: Dordrecht), 131

%\bibitem[Feast \& Whitelock(2000)]{fw00} Feast, M. \& Whitelock,
%P. 2000, in "The Evolution of the Milky Way: stars versus clusters,"
%edited by F. Matteucci \& F. Giovannelli (Kluwer: Dordrecht), 229
\bibitem[Feast \& Whitelock(2000)]{fw00} Feast, M. \& Whitelock,
P. 2000, in ``The Evolution of the Milky Way: stars versus clusters,''
eds. Matteucci \& Giovannelli, (Kluwer: Dordrecht), 229

\bibitem[Fourcade \& Laborde(1966)]{fourcade66} Fourcade, C. R. \&
Laborde, J. R. 1966, ``Atlas y Catalogo de Estrellas Variables en
C\'{u}mulos Globulares al Sur de -29\degr,'' C\'{o}rdoba

\bibitem[Frogel \& Whitelock(1998)]{frogel98} Frogel, J. A. \&
Whitelock, P. A. 1998, \aj, 116, 754

\bibitem[Girardi et al.(2002)]{girardi02} Girardi, L., Bertelli, G.,
Bressan, A., Chiosi, C., Greonewegen, M. A. T., Marigo, P., Salasnich,
B., \& Weiss, A. 2002, \aap, 391, 195

\bibitem[Harris(1996)]{harris96} Harris, W. E. 1996, \aj, 112, 1487

\bibitem[Ita et al.(2004)]{ita04} Ita, Y., Tanab\'{e}, T., Matsunaga,
N. et al. 2004, \mnras, 347, 720

\bibitem[Layden {\it et al.}(1999)]{layden99} Layden, A. C., Ritter,
L. A., Welch, D. W., \& Webb, T. M. A. 1999, \aj, 117, 1313

%\bibitem[Layden \& Sarajedini(2000)]{layden00} Layden, A. C. \&
%Sarajedini, A. 2000, \aj, 119, 1760

\bibitem[Layden {\it et al.}(2003)]{layden03} Layden, A. C., Bowes, B.
T., \& Webb, T. M. A. 2003, \aj, 126, 255

\bibitem[Lebzelter \& Wood(2005)]{lebwood05} Lebzelter, T. \& Wood,
P. R. 2005, \aap, 441, 1117

%\bibitem[Liller(1975)]{liller75} Liller, M. H. 1975, \apjl, 201, 125

%\bibitem[Lloyd Evans \& Menzies(1973)]{lem73} Lloyd Evans, T. \&
%Menzies, J. R. 1973, IAU Colloq. 21, edited by J. D. Fernie, (Reidel: ?????), 151

%\bibitem[Nemec, Nemec, \& Lutz(1994)]{nemec94} Nemec, J. M., Nemec, A.
%F. \& Lutz, T. E. 1994, \aj, 108, 222

\bibitem[Nysewander et al.(2009)]{nysewander09} Nysewander, M.,
Reichart, D. E., Crain, J. A., Foster, A., Haislip, J., Ivarsen, K.,
Lacluyze, A., \& Trotter, A. 2009, \apj, 693, 1417

\bibitem[Olivier \& Wood(2005)]{olivwood05} Olivier, E. A. \& Wood,
P. R. 2005, \mnras, 362, 1396

\bibitem[Percy et al.(2004)]{percy04} Percy, J. R., Bakos, A. G.,
Besla, G., Hou, D., Velocci, V. \& Henry, G. W. 2004, in ASP
Conf. Ser. 310, ``Variable Stars in the Local Group,'' eds. Kurtz \&
Pollard, (ASP: San Francisco), 348

% Moes intro
\bibitem[Piotto et al.(2007)]{piotto2007} Piotto, G., Bedin, L.~R.,
Anderson, J., et al.\ 2007, \apjl, 661, L53

\bibitem[Pojmanski \& Maciejewski(2004)]{asas} Pojmanski, G. \&
Maciejewski, G. 2004, Acta Astronomica, 54, 153

\bibitem[Reichart et al.(2005)]{reichart05} Reichart, D., Nysewander,
M. Moran, J. et al. 2005, Il Nuovo Cimento C, 28, 767

\bibitem[Richtler et al.(1994)]{richtler94} Richtler, T., Grebel,
E. K., \& Segewiss, W. 1994, \aap, 290, 412

\bibitem[Richtler(1995)]{richtler95} Richtler, T. 1995, \aaps, 109, 1

\bibitem[Robin et al.(2003)]{robin03} Robin, A. C., Reyl\'{e}, C.,
Derri\'{e}re, S. \& Picaud, S. 2003, \aap, 409, 523
%Reylé, S., Derrière and S. Picaud. A synthetic view on structure and evolution of the Milky Way

\bibitem[Sarajedini \& Norris(1994)]{saraj94} Sarajedini, Ata \& Norris,
J. E. 1994, \apjs, 93, 161

\bibitem[Schlafly \& Finkbeiner(2011)]{schlafly11} Schlafly, E. F. \&
Finkbeiner, D. P. 2011, \apj, 737, 103

% Moes intro
\bibitem[Smith(1995)]{smith1995} Smith, H.~A.\ 1995, ``RR Lyrae
Stars,'' (Cambridge Univ. Press: Cambridge)

\bibitem[Stellingwerf(1978)]{stellingwerf78} Stellingwerf, R.~F. 1978,
\apj, 224, 953

%\bibitem[Stetson(1981)]{stetson81}  Stetson, P. B. 1981, \aj, 86 687

\bibitem[Stetson(1987)]{stetson87}  Stetson, P. B. 1987, \pasp, 99, 191

\bibitem[Stetson(1994)]{stetson94}  Stetson, P. B. 1994, \pasp, 106, 250

\bibitem[Stetson(2000)]{stetson00}  Stetson, P. B. 2000, \pasp, 112, 925

% (31 other authors) 
\bibitem[Strutskie et al.(2006)]{strutskie06} Strutskie, M. F. et
al. 2006, \aj, 131, 1163

%\bibitem[Sumerel et al.(2004)]{sumerel04} Sumerel, A. N., Corwin,
%T. M., Catelan, M., Borissova, J., \& Smith, H. 2004, IBVS, 5533

%Moe's intro
\bibitem[Sweigart \& Catelan(1998)]{sweigar1998} Sweigart, A.~V., \&
Catelan, M.\ 1998, \apjl, 501, L63

%\bibitem[Walker(1998)]{walker98}  Walker, A. R. 1998, \aj, 116, 220

%\bibitem[Wehlau et al.(1978)]{wehlau78}  Wehlau, A., Liller, M. H., 
%Demers, S., \& Clement, C. C. 1978, \aj, 83, 598

%\bibitem[Wehlau et al.(1982)]{wehlau82} Wehlau, A., Liller, M. H.,
%Clement, C. C. \& Wizinowich, P. 1982, \aj, 87, 1295

\bibitem[Wood et al.(1999)]{wood99} Wood, P. R., Alcock, C. et
al. 1999, in IAU Symp. 191, ``Asymptotic Giant Branch Stars,'' eds. Le
Bertre, Lebre, \& Waelkens, 151

\bibitem[Wood, Olivier \& Kawaler(2004)]{wood04} Wood,
P. R., Olivier, E. A. \& Kawaler, S. D. 2004, \apj, 604, 800

\bibitem[Zinn \& West(1984)]{zinnwest84} Zinn, R. \& West, M. J. 1984,
\apjs, 55, 45

\bibitem[Zinn(1985)]{zinn85} Zinn, R. 1985, \apj, 293, 424

%\bibitem[Zorotovic et al.(2010)]{zorotovic} Zorotovic, M., Catelan, 
%M., Smith, H. A., et al. 2010, \aj, 139, 357
 

\end{thebibliography}
\end{document}